# The Atacama Cosmology Telescope: Systematic Transient Search of Single Observation Maps

Emily Biermann[1,2], Yaqiong Li[3,4], Sigurd Naess[5], Steve K. Choi[6], Susan E. Clark[7,8], Mark Devlin[9], Jo Dunkley[10,11], P. A. Gallardo[12], Yilun Guan[13], Allen Foster[14], Matthew Hasselfield[15], Carlos Hervías-Caimapo[16], Matt Hilton[17,18], Adam D. Hincks[19,20], Anna Y. Q. Ho[21], John C. Hood, II[22], Kevin M. Huffenberger[23,24], Arthur Kosowsky[1], Michael D. Niemack[3,21], John Orlowski-Scherer[9], Lyman Page[10], Bruce Partridge[25], Maria Salatino[7,8], Cristóbal Sifón[26], Suzanne T. Staggs[10], Cristian Vargas[16], and Edward J. Wollack[27]

[1] Department of Physics and Astronomy, University of Pittsburgh, Pittsburgh, PA 15213, USA; ebiermann@lanl.gov
[2] Electromagnetic Sciences & Cognitive Space Applications (ISR-2), Los Alamos National Laboratory, Los Alamos, NM 87544, USA
[3] Department of Physics, Cornell University, Ithaca, NY 14853, USA
[4] Key Laboratory of Particle Astrophysics, The Institute of High Energy Physics, Beijing, 100049, People's Republic of China
[5] Institute of Theoretical Astrophysics, University of Oslo, Norway
[6] Department of Physics and Astronomy, University of California, Riverside, CA 92521, USA
[7] Department of Physics, Stanford University, Stanford, CA 94305, USA
[8] Kavli Institute for Particle Astrophysics & Cosmology, P.O. Box 2450, Stanford University, Stanford, CA 94305, USA
[9] Department of Physics and Astronomy, University of Pennsylvania, 209 S. 33rd Street, Philadelphia, PA 19104, USA
[10] Joseph Henry Laboratories of Physics, Jadwin Hall, Princeton University, Princeton, NJ 08544, USA
[11] Department of Astrophysical Sciences, Peyton Hall, Princeton University, Princeton, NJ 08544, USA
[12] Kavli Institute for Cosmological Physics, University of Chicago, Chicago, IL 60637, USA
[13] Dunlap Institute for Astronomy and Astrophysics, University of Toronto, 50 St. George Street, Toronto, ON M5S 3H4, Canada
[14] Department of Physics, Princeton University, Jadwin Hall, Princeton, NJ 08540, USA
[15] Center for Computational Astrophysics, Flatiron Institute, 162 5th Avenue 9th floor, New York, NY 10010, USA
[16] Instituto de Astrofísica and Centro de Astro-Ingeniería, Facultad de Física, Pontificia Universidad Católica de Chile, Av. Vicuña Mackenna 4860, 7820436 Macul, Santiago, Chile
[17] Wits Centre for Astrophysics, School of Physics, University of the Witwatersrand, Private Bag 3, 2050, Johannesburg, South Africa
[18] Astrophysics Research Centre, School of Mathematics, Statistics, and Computer Science, University of KwaZulu-Natal, Westville Campus, Durban 4041, South Africa
[19] David A. Dunlap Department of Astronomy & Astrophysics, University of Toronto, 50 St. George Street, Toronto, ON M5S 3H4, Canada
[20] Specola Vaticana (Vatican Observatory), V-00120, Vatican City State
[21] Department of Astronomy, Cornell University, Ithaca, NY 14853, USA
[22] Department of Astronomy and Astrophysics, University of Chicago, 5640 South Ellis Avenue, Chicago, IL 60637, USA
[23] Florida State University, Tallahassee, FL 32306, USA
[24] Texas A&M University, College Station, TX 77843, USA
[25] Astronomy Department, Haverford College, 370 Lancaster Avenue, Haverford, PA 19041, USA
[26] Instituto de Física, Pontificia Universidad Católica de Valparaíso, Casilla 4059, Valparaíso, Chile
[27] NASA Goddard Space Flight Center, 8800 Greenbelt Road, Greenbelt MD 20771, USA

## Abstract

We conduct a systematic search for astrophysical transients using data from the Atacama Cosmology Telescope. The data were taken from 2017 to 2022 in three frequency bands spanning 77 to 277 GHz. In this paper, we present a pipeline for transient detection using single-observation maps where each pixel of a map contains one observation with an integration time of approximately 4 minutes. We detect 34 transient events at 27 unique locations. All but two of the transients are associated with Galactic stars and exhibit a wide range of properties. We also detect an event coincident with the classical nova YZ Ret and one event consistent with a flaring active galactic nucleus. We notably do not detect any reverse shock emission from gamma-ray bursts, a nondetection that may be in tension with current models.

## 1. Introduction

Multiwavelength studies of transient events are vital to understanding their underlying physical mechanisms. Until recently, millimeter transient science was limited to targeted follow-up observations. This gap is quickly being filled by harnessing surveys of the cosmic microwave background (CMB) to perform blind transient searches in millimeter wavelengths (N. Whitehorn et al. 2016; S. Guns et al. 2021; S. Naess et al. 2021a; Y. Li et al. 2023; C. Tandoi et al. 2024). Observations in the millimeter and radio allow us to probe nonthermal emission from shocks and jets (P. Chandra 2016; R. A. Chevalier & C. Fransson 2017) or magnetic reconnection in the atmospheres of nearby magnetically active stars (M. A. MacGregor et al. 2021). T. Eftekhari et al. (2022) predicted event rates of extragalactic synchrotron transients emitting in the millimeter for CMB surveys such as CMB-S4 (K. Abazajian et al. 2022), Simons Observatory (SO; P. Ade et al. 2019), The Atacama Cosmology Telescope (ACT; J. W. Fowler et al. 2007; R. J. Thornton et al. 2016), and the South Pole Telescope (SPT; J. E. Carlstrom et al. 2011). In particular, they found CMB surveys will serve as unbiased probes of the prevalence of reverse shocks (RSs) within long gamma-ray bursts (LGRBs), a phenomenon that occurs when two shock waves collide after the burst. Unlike targeted observations, blind searches for these events can provide constraints on the prevalence of RSs within gamma-ray bursts (GRBs; T. Eftekhari et al. 2022).

The list of high-energy extragalactic transient detections within millimeter wave bands is rapidly growing. This list includes observations of GRB afterglows (e.g., E. Berger et al. 2003; N. Kuno et al. 2004) and polarized RSs (T. Laskar et al. 2019), as well as tidal disruption events (e.g., E. Berger et al. 2012; Y. Cendes et al. 2021, Y. Yao et al. 2024). High-sensitivity wide-field CMB surveys are expected to add to this list. T. Eftekhari et al. (2022) predict that the ACT should observe two to 10 RS emission events from GRBs and that SO may observe dozens. The millimeter band is particularly suited to detecting early signatures of RS observations because the millimeter emission peaks only a couple hours after the trigger (R. Sari et al. 1998; J. S. Bright et al. 2023). T. Eftekhari et al. (2022) also predict a small chance of detecting tidal disruption events that have also been observed in targeted millimeter wavelength campaigns (Q. Yuan et al. 2016). We may also see emission from events similar to the extragalactic transient AT2018cow (S. J. Prentice et al. 2018), an unprecedented millimeter transient (A. Y. Q. Ho et al. 2019) that may have been a supernova or tidal disruption event (D. A. Perley et al. 2018; L. E. Rivera Sandoval et al. 2018; R. Margutti et al. 2019).

With a zoo of transients to search for, wide-field millimeter surveys designed for mapping the CMB are increasingly being used for transient detection. N. Whitehorn et al. (2016), using data from SPT, published the first blind transient search of CMB survey data and reported an event broadly consistent with a GRB afterglow with a peak flux of 16.5 ± 2.4 mJy at 150 GHz but with a low statistical significance due to a large number of trials ($p = 0.01$). S. Guns et al. (2021) used a single year of SPT-3G (an upgraded version of SPT) data to detect 13 stellar flares and two events consistent with flares from active galactic nuclei (AGN). Recently, C. Tandoi et al. (2024) detected 111 stellar flares from 66 stars over 4 yr and 1500 deg$^2$ of observations also using data from SPT-3G.

The ACT collaboration previously published three papers related to transient phenomena. Using 3 day coadded maps made for a Planet 9 search (S. Naess et al. 2021b), S. Naess et al. (2021a) serendipitously discovered three bright transients consistent with flares from magnetically active stars. This paper inspired a systematic transient search using the 3 day maps, yielding 17 transient detections, including those described in S. Naess et al. (2021a), mostly consistent with stellar flares. C. Hervías-Caimapo et al. (2024) also used ACT data to publish a targeted transient search putting upper flux limits on known tidal disruption events, supernovae, and GRBs.

In this paper, we expand upon Y. Li et al. (2023) with newly processed data using an improved pipeline on single-observation "depth-1" maps (defined more fully in Section 2). This pipeline searches for "bona fide transients," meaning objects that otherwise do not appear in the map, rather than variable sources. The maps used in this analysis are released publicly in ACT's recent Data Release 6 (DR6; S. Naess et al. 2025). The techniques described here prototype methods planned for a real-time transient pipeline for the Simons Observatory Large Aperture Telescope (SO-LAT; S. Parshley et al. 2018; N. Zhu et al. 2021), a CMB telescope being built in the Atacama Desert (P. Ade et al. 2019) with a field of view of 7°.8 at around 90 GHz. The SO-LAT will produce daily maps for transient searches (SO Collaboration 2024, in preparation).

This paper is organized as follows. In Section 2, we describe the ACT and the maps used for the analysis. In Section 3, we outline the transient detection pipeline itself. Then, in Section 4, we summarize the results and the performance of the pipeline. Finally, in Section 5, we discuss the origins of the transients, compare our findings to theoretical predictions, and comment on the future of CMB survey transient detection by predicting event rates for the upcoming SO.

**Table 1**
The Frequency Band, Observation Time Period, and FWHM Beam Size for Each Detector Array in AdvACT

| Array | Freq. Band | Op. Time | Beam (arcmin) |
|---|---|---|---|
| pa4 | f220 and f150 | 2016 Aug–2022 Oct | 1.09, 1.54 |
| pa5 | f150 and f090 | 2017 Apr–2022 Oct | 1.51, 2.24 |
| pa6 | f150 and f090 | 2017 Apr–2020 Jan | 1.55, 2.27 |
| pa7 | f040 and f030 | 2020 Feb–2022 Oct | 5.41, 7.26 |

**Note.** The beam size is given for each frequency band listed respectively. In 2020, pa7 replaced pa6 in the third optics tube, but the data of pa7 are not considered in this paper since the data are not ready for analysis.

## 2. Data

### 2.1. The Atacama Cosmology Telescope

The ACT was a 6 m off-axis Gregorian telescope located in northern Chile (J. W. Fowler et al. 2007; R. J. Thornton et al. 2016). In this paper, we use the data taken by the third generation of the ACT receiver, Advanced ACTPol (AdvACT), which simultaneously housed three optics tubes (S. W. Henderson et al. 2016), each containing a single dichroic, polarization-sensitive detector array (S.-P. P. Ho et al. 2017; S. K. Choi et al. 2018). The array frequencies and dates on the sky are summarized in Table 1. The data set we analyze was taken from 2017 to 2022 and covers three frequency bands: f220 (182–277 GHz), f150 (124–172 GHz), and f090 (77–112 GHz) with approximate full-width half-maximum (FWHM) beam sizes of 1′, 1′.4, and 2′.0 respectively (R. J. Thornton et al. 2016). The polarization-sensitive arrays of detectors in each tube are referred to as "pa*N*," where *N* is an array index. Since each array is dichroic, we typically refer to combinations of arrays and frequency bands, called "array bands," such as pa4-f150 or pa4-f220.

### 2.2. Depth-1 Maps

During the AdvACT survey, ACT surveyed the sky by performing wide constant-elevation scans (approximately 54°–94° peak to peak in azimuth; S. K. Choi et al. 2020) for hours at a time while the sky drifted past. During each scan, each point on the sky rotated into the elevation band covered by the detector arrays and was observed a few times by a subset of the detectors due to the scanning motion, before rotating out of that band, all over the course of a few minutes. Once done scanning, the telescope repositioned and started a new constant-elevation scan elsewhere. As part of DR6, we mapped each constant-elevation scan into a separate 0.′5-resolution map, which we call "depth-1 maps," since each sky pixel has drifted across the focal plane only once, in contrast to typical cosmology maps, which consist of coadditions of many days' observations. The depth-1 maps vary greatly in size, but 1500 $\deg^2$ is typical, with a pixel size of approximately a half arcminute.

These maps are described in detail in S. Naess et al. (2025). To summarize, these are maximum-likelihood maps using a standard nearest-neighbor pointing matrix. To save computing resources, we stopped the conjugate gradient solution process after 100 iterations, which suppresses power on scales larger than 10′, or, in spherical harmonic space, on multipoles $\ell \lesssim 1000$. These scales are irrelevant for the point-source-like events we consider in this study.

In contrast to the 3 day map analysis in Y. Li et al. (2023), the depth-1 maps do not go through a mean subtraction process since we find the negative ring surrounding the bright point sources causes false detections when the point sources' flux densities decrease. Therefore the depth-1 maps $\hat{m}$ (in temperature units) are directly matched and filtered to optimize point-source detection, giving an estimate of the flux density $F$ and signal-to-noise ratio (S/N) in each pixel of

$$F = \frac{\rho}{\kappa} = \frac{B^T U^{-1}\hat{m}}{\mathrm{diag}(B^T U^{-1} B)} \tag{1}$$

$$\mathrm{S/N} = \frac{\rho}{\sqrt{\kappa}} = \frac{B^T U^{-1}\hat{m}}{\sqrt{\mathrm{diag}(B^T U^{-1} B)}}, \tag{2}$$

where $\rho$ and $\kappa$ are the inverse-variance-weighted flux density and the inverse variance; $B$ includes the instrument beam and a factor converting flux density to a temperature unit; and $U$ is the noise covariance matrix of $\hat{m}$, which for point-source detection includes instrumental and atmospheric noise, clusters, and the CMB signal. Here we use $U^{-1} \approx HC^{-1}H$, where $H$ is a diagonal matrix with the diagonal representing the square root of the white-noise inverse variance map $\omega$ of $\hat{m}$, and $C$ is constructed directly from the power spectrum of the whitened map, $\omega^{0.5} * \hat{m}$, with the assumption that the noise in each frequency bin is independent. Each depth-1 map is also accompanied by a cosampled time map containing information about when each sky pixel was observed by the array.

There are some depth-1 maps with known pointing errors because there are few or no known point-source positions to calibrate the map position. Although the entire map may contain some point sources, the calibration is done on approximately 10 minute timescales. Therefore, these "bad pointing" maps contain regions where the pointing is off and are removed from the analysis. These maps make up about 5% of the total data set.[28]

## 3. Methods

### 3.1. Detection Pipeline

After applying the filtering process to produce the $F$ and S/N maps, we mask each depth-1 map. First, we mask dust measured from Planck and the Galactic plane by masking $-5^\circ < b < 5^\circ$ and declinations above 21.°5 and below $-6^\circ$ for $b > 5^\circ$, and above 21.°0 and below $-61.5^\circ$ for $b < -5^\circ$.[29] We also mask areas of the map within 10 pixels (about 5′) of the edge, where the noise properties fluctuate. We place a 5′ radius around known ACT sources with an average depth-1 S/N > $1\sigma$ in each of the three bands (see Table 2). We use this flux limit instead of masking all sources so that we will detect transients from known sources that are nominally not detected in a depth-1 map. These sources are typically AGN and dusty star-forming galaxies. We also mask map areas within 50′ of planets and map areas within 30′ of the blazar 3C 454.3. Bright sources have ringing effects associated with filtering, so large areas of the map around them must be masked. These masks are summarized in Table 3.

**Table 2**
Typical Detection Level (the Average Flux That Produces S/N > 5) and the Minimum Flux of Masked Sources for Each Frequency

| Frequency | $5\sigma$ Detection Level (mJy) | Minimum Flux (mJy) |
| --- | --- | --- |
| f090 | 143 | 30 |
| f150 | 269 | 50 |
| f220 | 443 | 90 |

**Table 3**
The Percent of Pixels of All Depth-1 Maps Cut by Each Type of Mask

| Mask | Radius | Percent of Pixels Cut |
| --- | --- | --- |
| Galaxy | ⋯ | 6.7 |
| Edge | 0.167′ | 1.6 |
| Source catalog | 5′ | 0.34 |
| Planets | 50′ | 0.004 |
| Blazar 3C 454.3 | 30′ | 0.003 |
| Combined | ⋯ | 8.4 |

**Note.** The Galaxy mask covers the Galactic plane, and the edge cut masks pixels within 10 pixels of the edge. The source cut masks all pixels within 5′ of any source above our flux limit in the ACT source catalog. The planets mask cuts pixels within 50′ of Venus, Mars, Jupiter, Saturn, Uranus, and Neptune. The blazar cut masks sources within 30′ of the bright blazar 3C 454.3. We also provide the percentage of pixels cut by all the masks combined without accounting for overlap between masks.

After masking the depth-1 maps, we apply a renormalization step on the matched filtered S/N depth-1 maps described in Section 2. This step would not be necessary in ideal situations, under which the non-point-source S/N is approximately white and the filtered maps' S/N would by its construction be approximately a normal distribution with a unit variance. Indeed, the mean of the squared S/N estimated for each of our depth-1 maps from Equation (2) is approximately 1, but some show localized enhanced noise along the scanning direction, manifesting as localized high- or low-S/N patterns at the size of a few arcminutes. See the top panel of Figure 1 for an example. Possible causes for these effects include uneven scan coverage due to low hit counts and/or a temporary calibration failure (Y. Li et al. 2023). The deviation of the S/N from Equation (2) from a normal distribution arises because the filter described in Section 2 estimates each noise covariance matrix $U$ from the entirety of the corresponding depth-1 map. Thus, $U$ underestimates the S/N in high-quality regions and overestimates S/N in poor-quality data regions.

[28] We checked these maps for transients as a separate analysis and did not find any. They are formally cut from this analysis since we are interested in automating the process as much as possible for future implementations with SO transient detection.

[29] The source, dust, and Galactic plane maps will be described in more detail in an upcoming ACT paper.

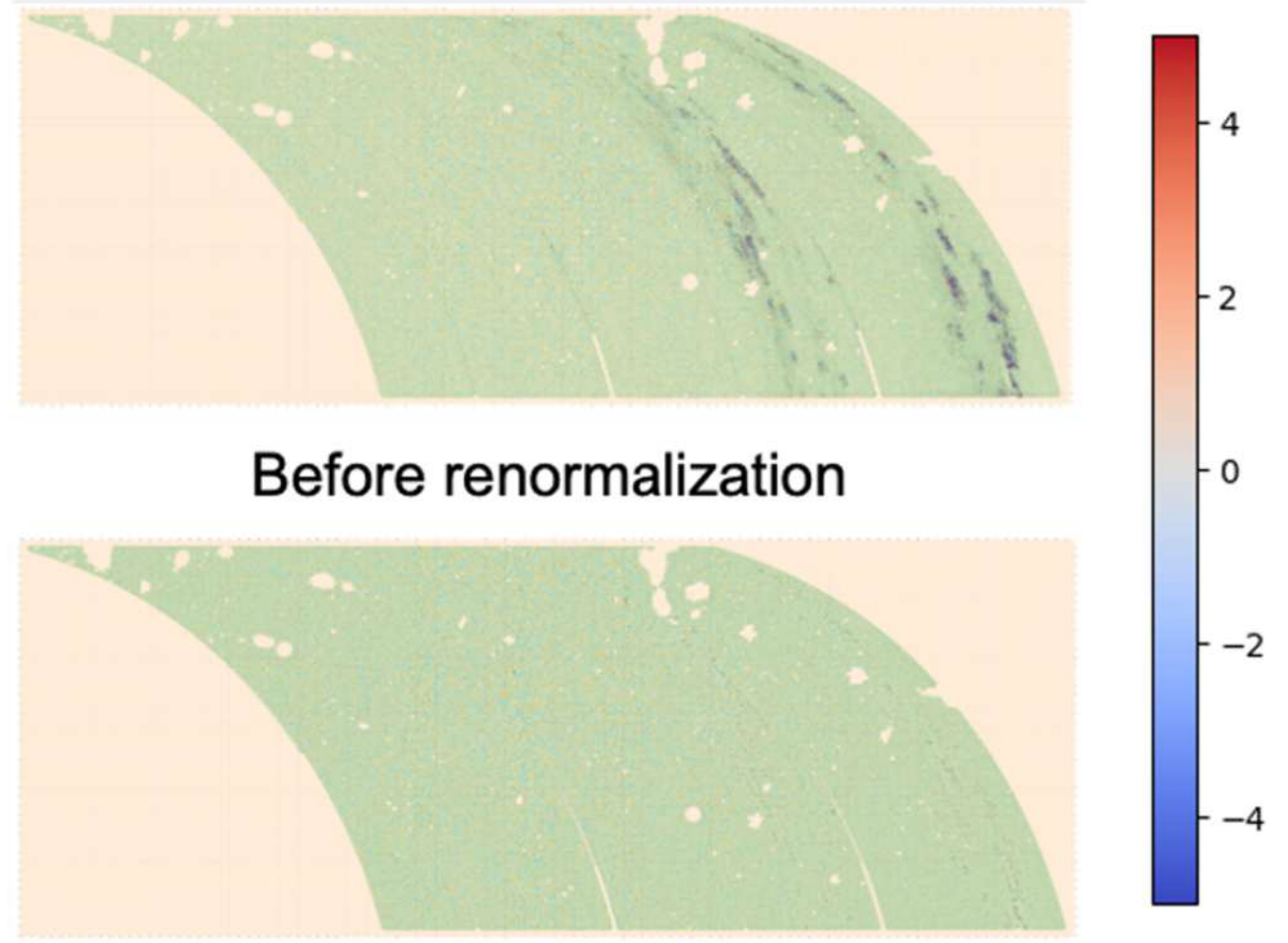

**Figure 1.** Example of S/N maps before and after renormalization, plotted on a linear color scale from −5 to +5. There are false signals showing high-S/N patches on the original matched filtered maps due to imprecise noise modeling. We cut the maps into 0°.5 × 0°.5 tiles and normalize each so that the mean of the squared S/N values for each tile is corrected to unity.

We compensate for the nonuniformity by dividing the maps into small tiles and renormalizing each tile. Specifically, each S/N map is split into approximately 0°.5 by 0°.5 tiles along the scanning direction of the telescope and corrected such that the mean of each tile is unity. The tile size is chosen to be small enough to be representative of small scales (10–20 times the beam size), but large enough so that the computation can be done quickly. To avoid including the signals from transients or other point sources, we calculate the normalization factor as the ratio of the median of the square of S/N in each tile to the median value of the square of a normal distribution value with $\sigma = 1$. See Figure 1.

Once the maps are renormalized, we perform our point-source detection in a similar way to Y. Li et al. (2023). First, we identify any pixels with an S/N greater than 5 and then use the "center of mass" evaluated by the flux of these pixels to obtain a position measurement. A key difference between this analysis and Y. Li et al. (2023) is the absence of mean subtraction. In this analysis, we detect all sources regardless of whether they are transient or stable and then cut the stable sources at a later step.

Even after S/N renormalization, many spurious detections still persist since the noise model of the matched filter does not account for localized noise patterns. We apply a second local matched filter around each candidate to correct for this. First, we cut a 1° by 1° thumbnail from the corresponding depth-1 unfiltered temperature map centered around the candidate's position. We then mask the center area and any existing point sources within the thumbnail. The cutting radius of both masks is twice of the beams' FWHM, and the masks and thumbnail's edge are also appodized with a 10′ radius. After applying the masks, we reapply a matched filter, using the power spectrum of the masked thumbnail to model the noise. After that, we repeat the source detection process on the matched filtered thumbnail. Detections that still have an S/N> 5 survive this cut.

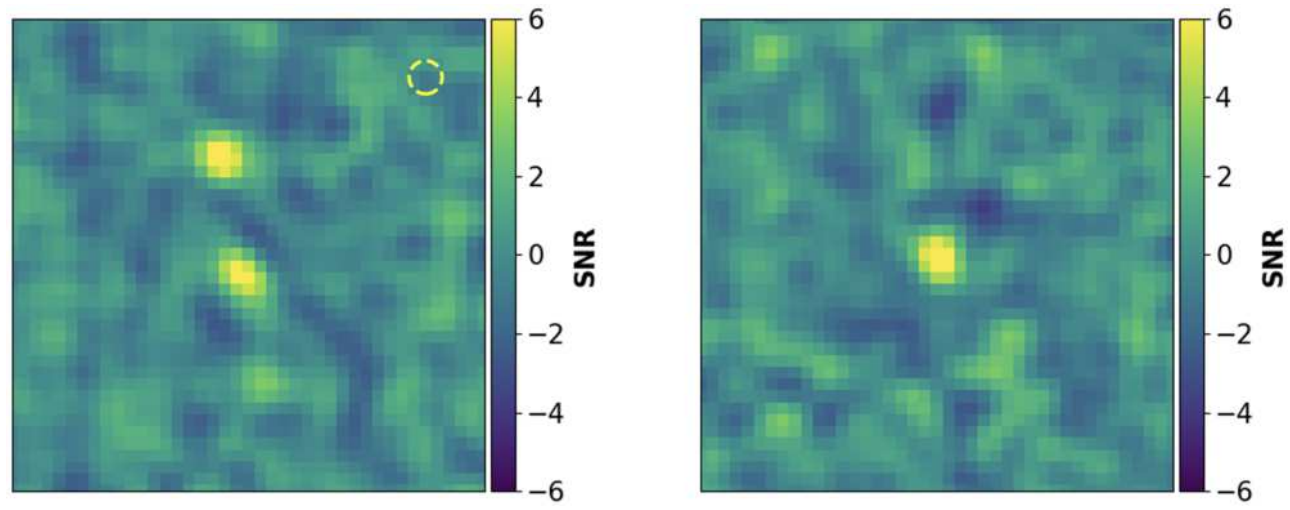

**Figure 2.** Example of a 20′ × 20′ map of an artifact (left) that passes all cuts and is manually cut from the analysis, in comparison with a map of a real candidate (right). This is an S/N plot made from a portion of a depth-1 map from array pa4-f150. For comparison, the dashed circle on the upper right corner shows the f150 beam size. If these sources are real transients, this would mean that two point sources separated by several arcminutes on the sky had a large rise in flux at the same time. The more likely scenario is that these are artifacts or fast moving objects such as satellites and not real transients.

Finally, when matching candidate sources in two or more maps, we require the candidate source positions to agree to better than 1′.5. Objects that appear in only one detector array are likely to be artifacts from glitches, so we require each candidate to appear in at least two arrays. We use a cross-matching radius a little larger than the beam size because some of the depth-1 maps have poor pointing and sources may be shifted.

One set of the depth-1 maps showed many obvious data artifacts so they were discarded.[30] The artifacts are likely due to poor observing conditions on that particular night.

In summary, we mask the Galactic plane, point sources, and planets; normalize the maps on degree scales; perform a matched filter on pixels with an S/N > 5; and delete one set of bad maps. After all of these cuts are applied, 524 candidate detections remain across the entire data set.

### 3.2. *Candidate Verification*

Some sources close to the flux cutoff of the ACT source catalog (see Table 2) are masked in some frequencies but not in others and appear as transient detections. To correct for this, we implement an additional cut on the candidates that requires the candidate not be within 3′ of a known source above the flux density limits in Table 2 in any frequency band regardless of which map the detection is made. This step cuts 112 out of the 524 detections.

Many of the remaining candidates are coincident with asteroids. These objects are treated in J. Orlowski-Scherer et al. (2024), so we cut them here using Astroquery's Skybot package (A. Ginsburg et al. 2019). Any candidate within 1′.5 of asteroids with a maximum $V$-band optical magnitude of 15 is cut. This cuts 358 out of the remaining 412 candidates.

Detections passing all cuts are confirmed visually using a diagnostic plot in which maps from each array band are plotted sequentially. A transient will not appear in the previous maps and should look like a point source when detected by the pipeline. We detect six artifacts that are cut from the transient list using this method. An example of an artifact is shown alongside a true transient detection in Figure 2. In this

[30] This corresponds to one map out of 5474 in pa5-f150, for example. Since we detect transients in 29 maps in this band, there is about a 0.5% chance this cut map contains a real transient.

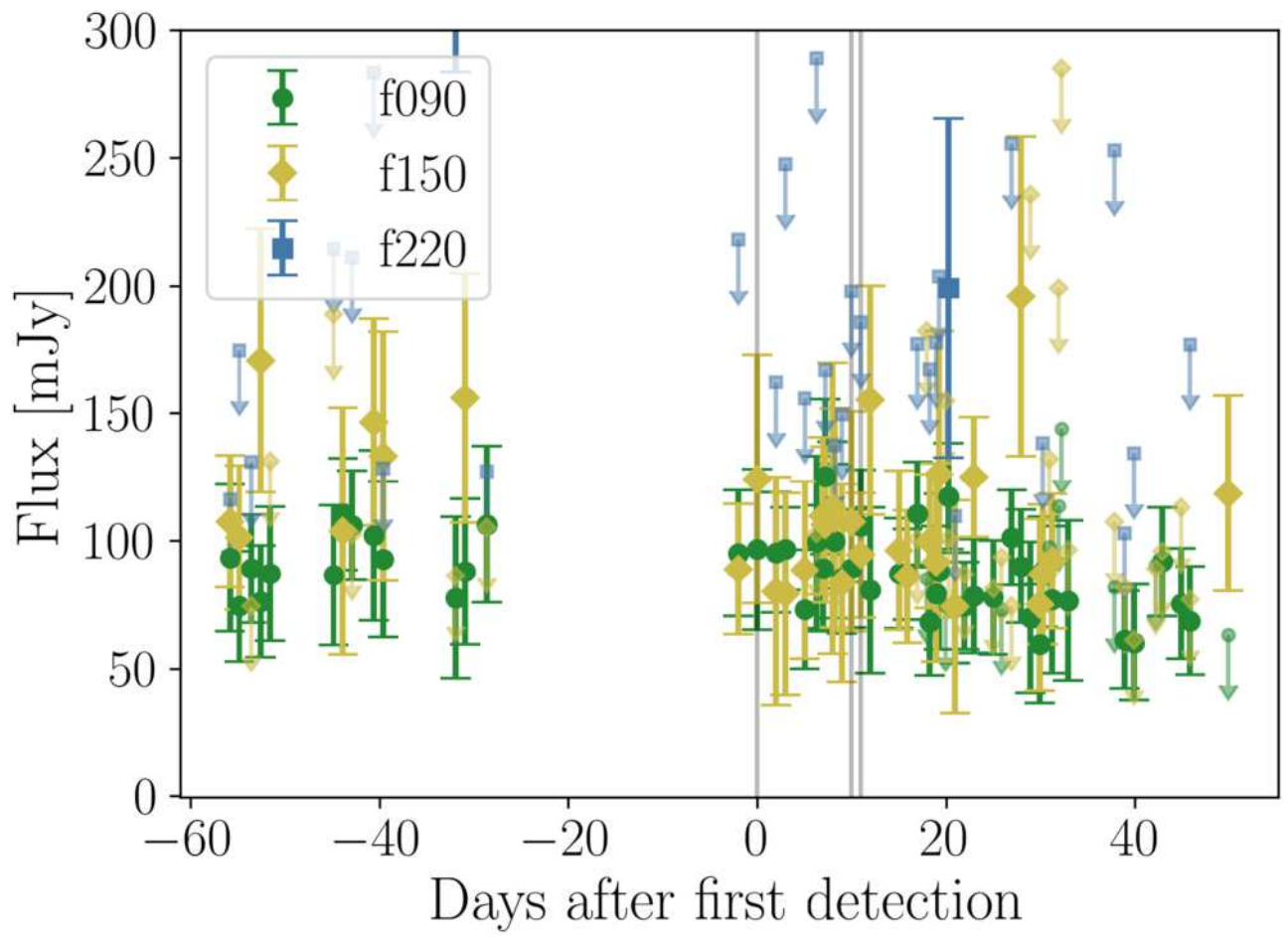

**Figure 3.** Event (R.A.: 9.063(6), decl.: −41.870(6)) detected by our transient pipeline but cut from the analysis since it is clearly a variable source rather than a transient event. This figure shows the light curve of this event. The gray vertical lines indicate where this source is detected by the pipeline; the data points with down arrows indicate upper limits (nondetections); and the data points with error bars indicate a $>3\sigma$ detection. Note there are many flux upper limits from f220 that are above the $y$-axis limit in this plot because this frequency band is substantially noisier than f090 or f150.

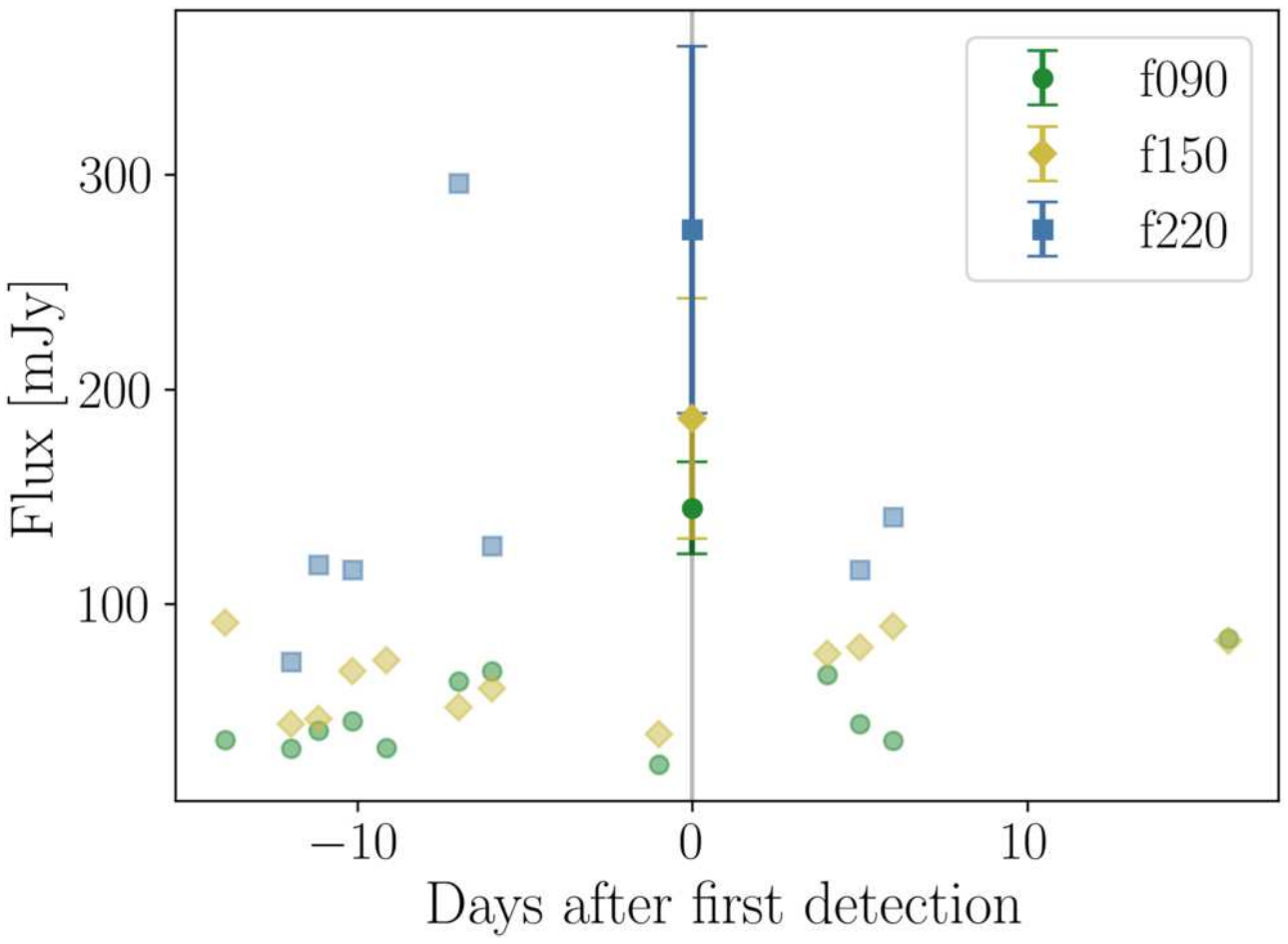

**Figure 4.** The light curve from the depth-1 maps of Event 1 (R.A.: −90.025 (4), decl.: 10.739(4)). The gray vertical lines indicate where this source is detected by the pipeline; the data points with down arrows indicate upper limits (nondetections); and the data points with error bars indicate a $>3\sigma$ detection. This serves as an example of a real transient detection. The light curves for all of the transients are shown in Figures A1 and A3.

**Table 4**
Summary of Candidate Cuts

| | Init. | MF | CM | AS | PS | MA | Rem. |
|---|---|---|---|---|---|---|---|
| pa4-f220 | 201682 | 150115 | 51290 | 265 | 0 | 3 | 9 |
| pa4-f150 | 113611 | 95025 | 18268 | 240 | 52 | 3 | 23 |
| pa5-f150 | 71963 | 58266 | 13249 | 339 | 77 | 3 | 29 |
| pa5-f090 | 84046 | 75314 | 8454 | 206 | 23 | 7 | 42 |
| pa6-f150 | 38723 | 33918 | 4574 | 161 | 53 | 1 | 16 |
| pa6-f090 | 59375 | 56107 | 3123 | 94 | 26 | 2 | 23 |

**Note.** We quote the total number of candidates detected in all of the depth-1 maps, the number of candidates that are cut at each step, and the number of remaining candidates. Init. and Rem. refer to the initial and remaining candidates, respectively. MF, CM, AS, PS, and MA refer to the number of candidates cut by the matched filter, array cross-matching, asteroid cross-matching, post-point-source cutting, and manual cutting, respectively.

instance, two point sources separated by several arcminutes appear to brighten at the same time, a phenomenon much more likely to be caused by an artifact in the data than a real transient event. The source of these artifacts is unknown, but they may be caused by glitches in the time stream due to cosmic rays or atmospheric variations.

We also cut three detections from a variable source at an R.A. 9.603(6)° and decl. −41.870(6)°. This object is a part of the ACT source flux catalog (ACT-S J0038.4-4152, R.A.: 9.60656 decl.: −41.87096), but the mean flux (24.31(0.72) mJy in f090 measured from the depth-1 maps) was below the source masking threshold. Figure 3 shows the light curve of the event, confirming this is a variable source and not a transient. The counterpart for this source is unclear because there is a high density of objects, but the closest match from the NASA/IPAC Extragalactic Databaseis WISEA J003825.68-415218.8, which is 3.″72 away. After this final cut, 45 detections remain, corresponding to 34 independent events.

Each step in the pipeline is summarized in Table 4. An example of a transient light curve (Event 1) is shown in Figure 4. We also count the number of sources with a single-observation S/N greater than 5 that are cut in the matched filter (MF) step as well as the array cross-matching (CM) step, as shown in Table 5. This allows us to determine the source-finding efficiency of our pipeline. The detection efficiency varies a lot between bands mostly due to the cross-matching step. Some bands perform very well, such as pa5-f090, and have comparatively little noise. In these bands, more candidates are detected but are cut because they are not detected in other, noisier, bands. Also, in some cases, the MF step cuts up to 22% of real sources. This is due to a suboptimal noise model, which could be optimized in future transient studies to produce better efficiency.

**Table 5**
Percentage of S/N > 5 Sources That Fail the Matched and Cross-matching Cuts

| | Init. | MF (%) | CM (%) | Total (%) |
|---|---|---|---|---|
| pa4-f220 | 30054 | 7.7 | 2.5 | 10.3 |
| pa4-f150 | 107729 | 12.0 | 2.4 | 14.4 |
| pa5-f150 | 180063 | 9.6 | 22.4 | 32.1 |
| pa5-f090 | 372672 | 15.4 | 36.6 | 51.9 |
| pa6-f150 | 110541 | 17.3 | 8.0 | 25.4 |
| pa6-f090 | 233834 | 22.1 | 21.9 | 43.9 |

**Note.** We present the number of such sources detected in all of the depth-1 maps (Init.), the percentage that are cut by applying a matched filter and requiring an S/N greater than 5 (MF), and by cross-matching across arrays (CM). We also give the combined number of detections cut by either step (Total).

## 4. Results

After all cuts are applied, 142 detections remain across all array bands corresponding to 45 unique detections or 34 unique transient events at 27 unique locations. There are fewer transient events than detections because some long-lived events are detected in more than one map. The events are summarized in Tables A1 and A2. A catalog of these transients is published alongside this paper. The columns of this csv file are outlined in Table A3.

In Figure A4, we include $10' \times 10'$ cutouts from sequential depth-1 maps centered on each event to further illustrate the transient nature of these objects. Each object is clearly missing from the map before detection and then remains, disappears for the rest of observations, or disappears and then reappears again at a later time. It is worth noting that in many cases the transients do not appear to flare in the f220 band. This is most likely because this band is noisier compared to f150 and f090 due to the atmosphere, and thus the flare has a lower S/N at this frequency.

We compare our work using single-observation maps to the transient detections made from the 3 day coadded maps in Y. Li et al. (2023); we recover 12 out of 21 detections. We do not recover any of the four detections, which the authors believe do not look like real transients (Events 2, 9, 10, and 13), indicating that they are indeed false detections. The true detections from Y. Li et al. (2023) that we do not recover in this paper are accounted for in the following ways. Three of the events from Y. Li et al. (2023), J060702+174157, J190222-53610, and J070038-111436, are masked in the depth-1 analysis as a different Galactic dust mask is used. One event, J060757-542626, is too dim to be detected in the depth-1 maps. Although we detect J225302+165027 b, J225302+165027 a is only seen in one array within the depth-1 maps (array 5), so it is cut from this analysis.

We search for counterparts for each transient using the SIMBAD database (M. Wenger et al. 2000).[31] Since the SIMBAD database mostly includes nearby bright stars, we calculate the probability of a chance association ($p$-value) by assuming a local density of Gaia stars (P. Gavras et al. 2023) with the counterpart's magnitude or lower, $\rho$, and a Poisson probability distribution modeling the chance of a random association,

$$p = 1 - \exp[-\pi \rho d^2], \tag{3}$$

where $d$ is the angular separation between the transient position and the counterpart position on the sky. For the events with Gaia counterparts (all but Event 26), we use the Gaia position to calculate the separation between the candidate and the counterpart. The results are summarized in Table A4. All of the $p$-values are less than $4.1 \times 10^{-4}$, indicating that the counterparts are likely correct associations.

We present light curves from the depth-1 maps in Figures A1 and A3. The flux of each depth-1 map is calibrated using light curves form Uranus (see C. Hervías-Caimapo et al. 2024) for a complete description of this calibration). Although the calibration factor is not yet published for f220, the calculation is the same. The depth-1 light-curve plots show measured detections with an S/N of at least 3 and give upper limits to all other data points. The upper limits are defined by calculating the 95% confidence interval for the positive part of the Gaussian distribution defined by the measured flux and the error on the flux. These light curves are published alongside this paper. The columns of these files are outlined in Table A5.

Most of the events are associated with flaring stars. The light curves from the depth-1 maps of these events are shown in Figure A1. Most of these events last less than a day and only include one detection from the depth-1 maps. In addition to these light curves, we provide high-resolution light curves made by binning detectors in each array into four groups, allowing us to study minute-scale flux deviations (Figure A2).

We detect two transients that cannot be reasonably associated with stellar flares (Events 22 and 26). Their light curves are shown in Figure A3. Event 22 is coincident with the classical nova YZ Ret, an optically bright and well-observed nova within our Galaxy. These observations are only the second millimeter observations of a classical nova. This event is discussed in Section 5 and will be further analyzed in a future paper.

Event 26 is associated with the LINER-type AGN 2MASX J19495127-3635239. Within a 2° radius around the transient's position there are 19 AGN within the Simbad catalog, giving this association a $p$-value of 0.06. If we only consider LINER-type AGN, this value drops to 0.003. There is a small chance this event is misclassified, but this is unlikely since the light curve shows similar behavior to other known AGN.

## 5. Discussion and Conclusions

### 5.1. Stellar Flares

The transients presented in this paper contribute to a growing list of radio and millimeter stellar flares. There are now several examples of millimeter and radio stellar flares from M dwarfs (M. A. MacGregor et al. 2018; A. M. MacGregor et al. 2020), RS CVn variables (A. J. Beasley & T. S. Bastian 1998; J. M. Brown & A. Brown 2006), T Tauri stars (G. C. Bower et al. 2003; M. Massi et al. 2006; D. M. Salter et al. 2010; S. Mairs et al. 2019), and more (S. Guns et al. 2021; S. Naess et al. 2021a; Y. Li et al. 2023; C. Tandoi et al. 2024). In Figure 5 (left) we provide a comparison of the luminosities of these events. The events exhibit a wide range of characteristics, indicating that a large parameter space of millimeter stellar flares exists. We notably do not detect any T Tauri stars as seen in G. C. Bower et al. (2003), M. Massi et al. (2006), D. M. Salter et al. (2010), and S. Mairs et al. (2019), which are expected to flare due to interactions with an accretion disk. All other papers listed present flares from K or M stars, but we also detect flares from six G-type stars and one ApSi star that are not found in millimeter stellar flare literature.

Since C. Tandoi et al. (2024) contains many events, we compare the distribution of spectral indices, which are evaluated by fitting a power law $S_\nu \propto \nu^\alpha$ of frequency $\nu$ using peak flux density $S_\nu$, in addition to luminosity in Figure 5 (right). The distributions are similar but ACT sees events with a slightly more negative spectral index. The combined distribution hints at a Gaussian distribution centered just above zero, indicating that the ACT and SPT stellar flare events are driven by similar mechanisms. Most of the events have a flat spectrum ($|\alpha| \lesssim 1$), which is often associated with gyrosynchrotron radiation, and two events have a falling spectrum (Events 5a, 15a, 15 c, $\alpha \lesssim 1$), which is associated with synchrotron emission. Synchrotron emission is similar to gyrosychrotron emission but with higher kinetic energies. Two of the stellar flares have rising spectra ($\alpha \gtrsim 0.5$, Events 2a and 18). It is possible that this is an indicator of different classes of stellar flares in M-type stars, but given the single peak in Figure 5, it is more likely that these are similar events with a wide distribution in intrinsic properties.

[31] https://simbad.cds.unistra.fr/simbad/

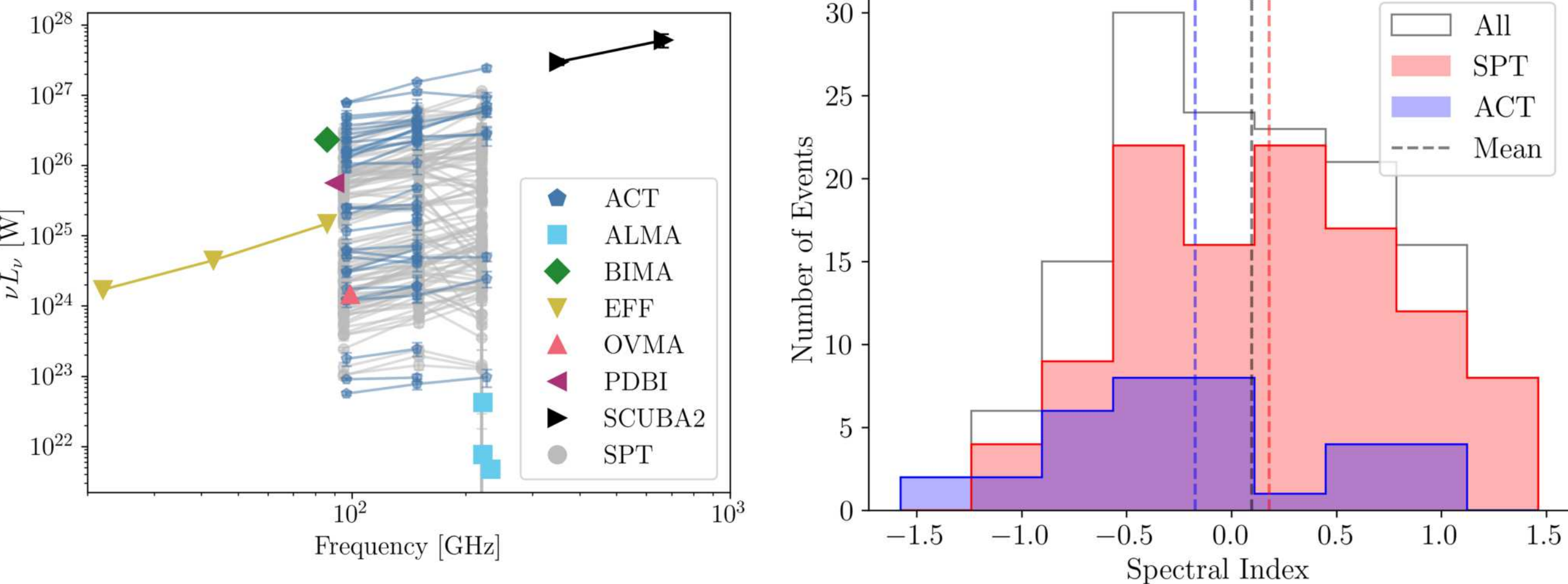

**Figure 5.** Left: the characteristic luminosity in each frequency band for stellar flare events. The luminosity is calculated using the distance to the flare's counterpart listed in Table A4. We include other millimeter flares for comparison from the Atacama Large Millimeter/submillimeter Array (ALMA; M. A. MacGregor et al. 2018; A. M. MacGregor et al. 2020), the Berkeley-Illinois-Maryland Association (BIMA; G. C. Bower et al. 2003), the Nobeyama 45 m telescope (EFF; T. Umemoto et al. 2009), the Owens Valley Millimeter Array (OVMA; J. M. Brown & A. Brown 2006), the Plateau de Bure Interferometer (PDBI; M. Massi et al. 2006), the JCMT Transient Survey (SCUBA2; S. Mairs et al. 2019), and the South Pole Telescope Flaring Stars Survey (SPT; C. Tandoi et al. 2024). Note that three of the events from C. Tandoi et al. (2024) do not have distance measurements and are not included in this figure. Right: histograms of spectral indices for stellar flares from this paper (ACT) and C. Tandoi et al. (2024; SPT). We also give the combined distribution. The means of the ACT, SPT, and combined distributions are $\alpha = -0.17 \pm 0.6$, $0.18 \pm 0.7$, and $0.10 \pm 0.6$, respectively.

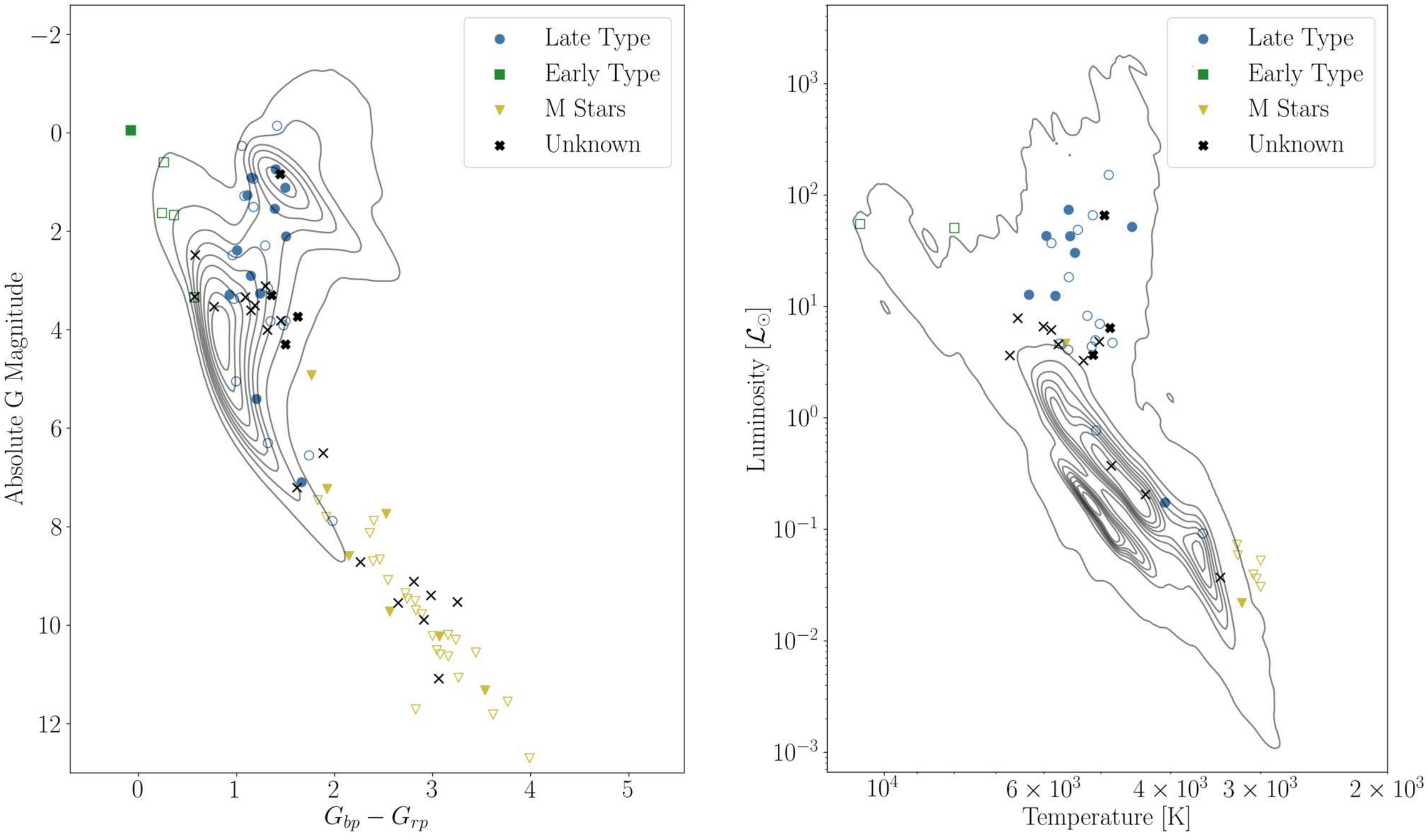

**Figure 6.** Color–magnitude diagram (left) and luminosity plot (right) of host stars of ACT (filled points) and SPT (unfilled points) transient detections. Luminosity is given in solar luminosity units, $\mathcal{L}_\odot$. The contours represent the density of Gaia stars within the ACT footprint. The host stars are separated into four categories: late type (O-F5), early type (F5-K), M-type stars (M and M dwarf), and unknown spectral type. The SPT and ACT populations appear to be consistent. The color, magnitude, temperature, and luminosity measurements of the host stars are all from the Gaia database with the exception of five SPT host-star distance measurements used to calculate the absolute G magnitude (see Table 2 in C. Tandoi et al. 2024). These measurements are instead taken from A. R. Riedel et al. (2014), F. van Leeuwen (2007), C. P. M. Bell et al. (2015), K. G. Stassun et al. (2019), and C. T. Finch et al. (2014). Note that some of the flares are missing from the plots since the measurements were missing from the Gaia database.

We also compare the ACT and SPT (C. Tandoi et al. 2024) transient host-star properties. Figure 6 shows a color–magnitude diagram (the host-star absolute magnitude as a function of the difference of two G-band magnitude measurements) and a luminosity plot (the host-star luminosity as a function of temperature). We split the host stars into four categories: late type (O-F5), early type (F5-K), M-type stars (M and M dwarf), and unknown spectral type. We find that the host stars seen by SPT and those seen by ACT are very similar to each other. Similarly to C. Tandoi et al. (2024), the M dwarfs' host stars in ACT lie above the main sequence and dominate the sample. Both samples detect flares from stars across spectral types, indicating millimeter flares, although more common in M dwarf stars, are present in early and late types as well.

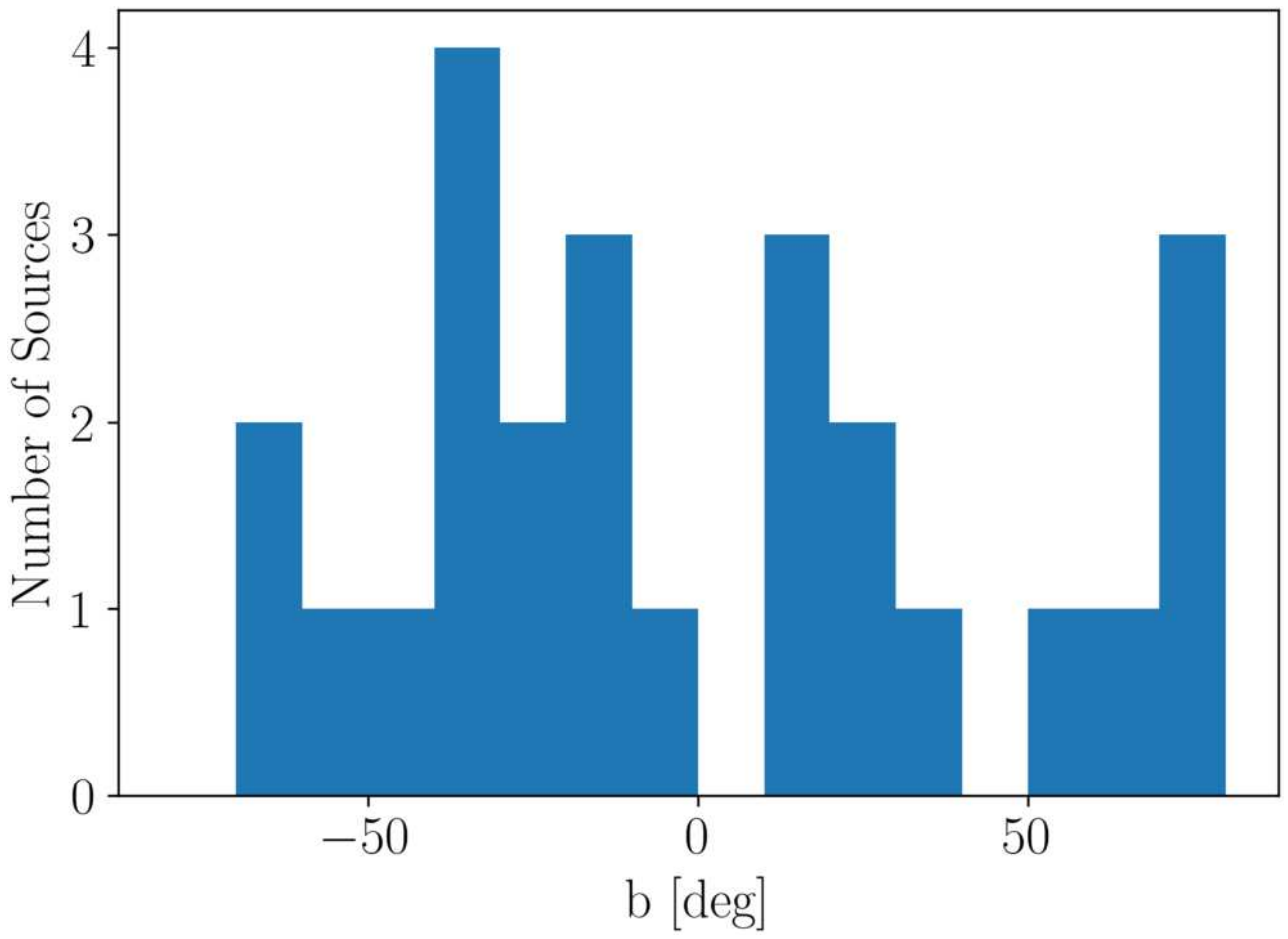

**Figure 7.** Histogram with a 10° bin size of the Galactic latitude of events that are associated with stars. This histogram hints at an even distribution; however, the number of flares is too small to make a broad generalization. Also, the sample is biased because the Galactic plane is masked.

Further analysis of these events must be done to uncover the mechanism behind these flares, which is beyond the scope of this paper. All of the stellar counterparts to the transient events we observe correspond to known magnetically active stellar types and may have similar emission mechanisms. Flares from nonthermal electrons typically occur due to optically thin gyrosynchrotron emission (J. M. Brown & A. Brown 2006; M. A. MacGregor et al. 2021). It has been proposed that bright stellar flares such as these could be driven by "elementary eruptive phenomenon" that involves the combination of energy from many heated flare loops (Z. Mouradian et al. 1983; M. A. MacGregor et al. 2021).

We also note that many of the stellar flares we detect are located close to the Galactic plane, as shown in Figure 7. This figure shows a histogram of the Galactic latitude of each stellar flare at a unique location. Although the number of stellar flares is too small to make definitive conclusions regarding this statistic, the transients seem to be evenly distributed corresponding to close and dim events. This result is expected since the Galactic plane ($b = 0$) is masked.

C. Tandoi et al. (2024) detect a factor of 10 more transients in their search of SPT data. The discrepancy is likely due to two factors: a different choice in the initial S/N detection threshold, and a difference in map properties. C. Tandoi et al. (2024) choose a $3\sigma$ S/N source detection versus the $5\sigma$ detection in this analysis and therefore detect a more complete sample of transients. The higher threshold used in this paper results in a higher purity at the expense of completion. Whereas many of the transients are filtered manually in C. Tandoi et al. (2024; the exact number is not provided in the paper), only nine detections, six from artifacts and three from a constant source, are cut manually in this analysis. Besides the S/N choice, SPT may detect more events since the noise in their maps are more Gaussian than the ACT depth-1 maps. The depth-1 maps are made from single array observations with noise properties that deviate from Gaussian distributions. This is addressed in the transient detection analysis by not allowing events only detected by one array. The SPT analysis averages their maps across all wafers, resulting in much more Gaussian noise properties.

The difference in event count from this analysis and C. Tandoi et al. (2024) suggests there are many more transients at lower S/N within ACT data. Also, the Galactic plane is masked in this analysis. Further analysis of the Galactic plane region will probe a largely unexplored population of potential millimeter transients. Although real-time detection efforts may benefit from higher purity samples, future archival transient studies may consider lower detection thresholds to achieve higher completion.

### 5.2. *Other Transients*

Few submillimeter and millimeter observations of classical novae exist in the literature. R. J. Ivison et al. (1993) observed Nova Cygni 1992 in wavelengths ranging from 0.42 to 2mm at 66, 104, 224, 234, 357, and 358 days after the outburst, finding their data to be consistent with free–free emission from an optically thick nova but inconsistent with canonical radio models. The remnant of Nova V5668 Sgr (2015) was observed in millimeter frequencies by Atacama Large Millimeter/submillimeter Array (ALMA; M. P. Diaz et al. 2018). These observations resolved the structure of the nova but were taken years after the actual explosion. The observations from Event 22 presented in this paper associated with YZ Ret are therefore only the second millimeter observations of a classical novae taken during the outburst.

The classical nova YZ Ret is a well-studied object with several X-ray and optical observations. The nova was first discovered on 2020 July 15.590 UT (E. O. Waagen 2020). This discovery is particularly notable because it is the first nova in which the X-ray flash phase was observed before it was detected in the optical (O. König et al. 2022). Our observations are coincident with X-ray observations that occurred about 60 days after the X-ray flash (K. Sokolovsky et al. 2020; K. V. Sokolovsky et al. 2022). These observations are associated with the supersoft X-ray source phase in which optically thick winds stop and X-ray emission is allowed to escape (I. Hachisu & M. Kato 2023). This emission is expected to follow a blackbody (K. V. Sokolovsky et al. 2022) and, although we indeed measure a spectral index value consistent with thermal emission at the peak of the event, the spectrum of our millimeter observations flatten over time. Future work will explore this discrepancy in more detail.

Event 26, which is associated with an AGN, is reminiscent of the two extragalactic events detected by SPT (S. Guns et al. 2021). Those events were also longer in duration (I. Márquez et al. 2017) than a stellar flare and are consistent with bright flares from AGN. Curiously, the events from SPT have positive spectra that flatten over time (consistent with a newly formed jet), whereas we measure a falling spectrum throughout the flare. It is also possible that this observation is not a true transient, but rather an intrinsic variability. More observations are required to make this distinction.

### 5.3. *Nondetection Implications*

We do not detect any transient signatures consistent with GRBs. Using a high-energy model, T. Eftekhari et al. (2022)

predict that over 7 yrs of observations, with an approximately 3 day cadence and a map noise level of 17 mJy, ACT should have observed over 10 $\Gamma = 50$ and more than two $\Gamma = 200$ LGRB RS events with an S/N of 10 or greater. A low-energy model predicts that ACT observed less than one and less than two LGRB RS events for gamma factors of 200 and 50, respectively. These predictions assume that all on-axis LGRBs produce observable RS emission. Current targeted observations of LGRBs put an upper limit of RS emission at $\geqslant 30\%$ of all LGRBs (T. Eftekhari et al. 2022). Therefore, we may not expect to detect any low-energy LGRB RS emission with ACT but conservatively expect to see three events over ACT's entire observation period given the high-energy model. It is possible that these events are detected but are cut from the analysis. A detection level of 170 mJy is optimistic for depth-1 maps (see Table 2) and, although such events may be detected in f090, they are less likely to be seen in f150 or f220 and thus may be cut in the cross-matching step.

Although the nondetection of GRB RS can be explained in this paper, targeted searches of GRBs in ACT data also result in nondetections (C. Hervías-Caimapo et al. 2024). Moreover, SPT, which T. Eftekhari et al. (2022) predict will observe up to 40 high-energy GRB events, do not report observing these transients. Two plausible explanations are that the low-energy model is more accurate than the high-energy model, or a fewer number of GRBs produce observable RS emission than expected.

### 5.4. Future Prospects

The successor high-resolution microwave telescope to ACT and SPT is the Large Aperture Telescope (LAT) of SO (SO Collaboration 2024, in preparation), which will begin regular science operations at the same site as ACT in 2025. The SO-LAT has several advantages over ACT for detection of microwave transients, including more detectors on a given transient in each scan due to a larger focal plane array; more frequency bands; a larger sky region covered as a regular scan strategy; and more regular time-domain coverage. The microwave transients detected in this work, as well as those from SPT (J. E. Carlstrom et al. 2011), set the stage for the more formal transient detection effort that SO will pursue. Clearer answers to intriguing scientific questions, such as the fraction of GRBs that produce transients via the RS mechanism, await the next generation of microwave transient experiments.

## Acknowledgments

The authors would like to thank the reviewer whose comments significantly improved this work.

Support for ACT was through the U.S. National Science Foundation through awards AST-0408698, AST-0965625, and AST-1440226 for the ACT project, as well as awards PHY-0355328, PHY-0855887, and PHY-1214379. Funding was also provided by Princeton University, the University of Pennsylvania, and a Canada Foundation for Innovation (CFI) award to UBC. ACT operated in the Parque Astronómico Atacama in northern Chile under the auspices of the Agencia Nacional de Investigación y Desarrollo (ANID). The development of multichroic detectors and lenses was supported by NASA grant Nos. NNX13AE56G and NNX14AB58G. Detector research at NIST was supported by the NIST Innovations in Measurement Science program. Computing for ACT was performed using the Princeton Research Computing resources at Princeton University, the National Energy Research Scientific Computing Center (NERSC), and the Niagara supercomputer at the SciNet HPC Consortium. SciNet is funded by the CFI under the auspices of Compute Canada, the Government of Ontario, the Ontario Research Fund–Research Excellence, and the University of Toronto. We thank the Republic of Chile for hosting ACT in the northern Atacama, and the local indigenous Licanantay communities, whom we follow in observing and learning from the night sky.

This work has made use of data from the European Space Agency (ESA) mission Gaia (https://www.cosmos.esa.int/gaia), processed by the Gaia Data Processing and Analysis Consortium (DPAC; https://www.cosmos.esa.int/web/gaia/dpac/consortium). Funding for the DPAC has been provided by national institutions, in particular the institutions participating in the Gaia Multilateral Agreement.

C.H.C. acknowledges ANID FONDECYT Postdoc Fellowship 3220255 and BASAL CATA FB210003. M.H. acknowledges financial support from the National Research Foundation of South Africa. A.D.H. acknowledges support from the Sutton Family Chair in Science, Christianity and Cultures, from the Faculty of Arts and Science, University of Toronto, and from the Natural Sciences and Engineering Research Council of Canada (NSERC; RGPIN-2023-05014, DGECR-2023-00180). K.M.H. acknowledges support from NSF award 2206344. S.N. acknowledges this work was supported by a grant from the Simons Foundation (CCA 918271, PBL). C.S. acknowledges support from the Agencia Nacional de Investigación y Desarrollo (ANID) through Basal project FB210003.

*Facility:* ACT.

*Software:* Astropy (Astropy Collaboration et al. 2013, 2018, 2022), Astroquery (A. Ginsburg et al. 2019), pixell (https://github.com/simonsobs/pixell).

## Appendix
## Supplemental Tables and Figures

Figure A1 presents light curves from the depth-1 maps for each stellar flare transient. Figure A2 presents subarray stellar flare light curves from forced photometry measurements. Figure A3 shows the depth-1 light curves for non-stellar-flare transients. Figure A4 depicts three-color images of depth-1 maps close to each event. Tables A1 and A2 present transient positions and characteristics, respectively. Table A3 gives column descriptions for the Depth-1 Transient Catalog accompanying this paper. Table A4 presents all transient counterparts. Finally, Table A5 gives column descriptions for the machine-readable depth-1 light curve files accompanying this paper.

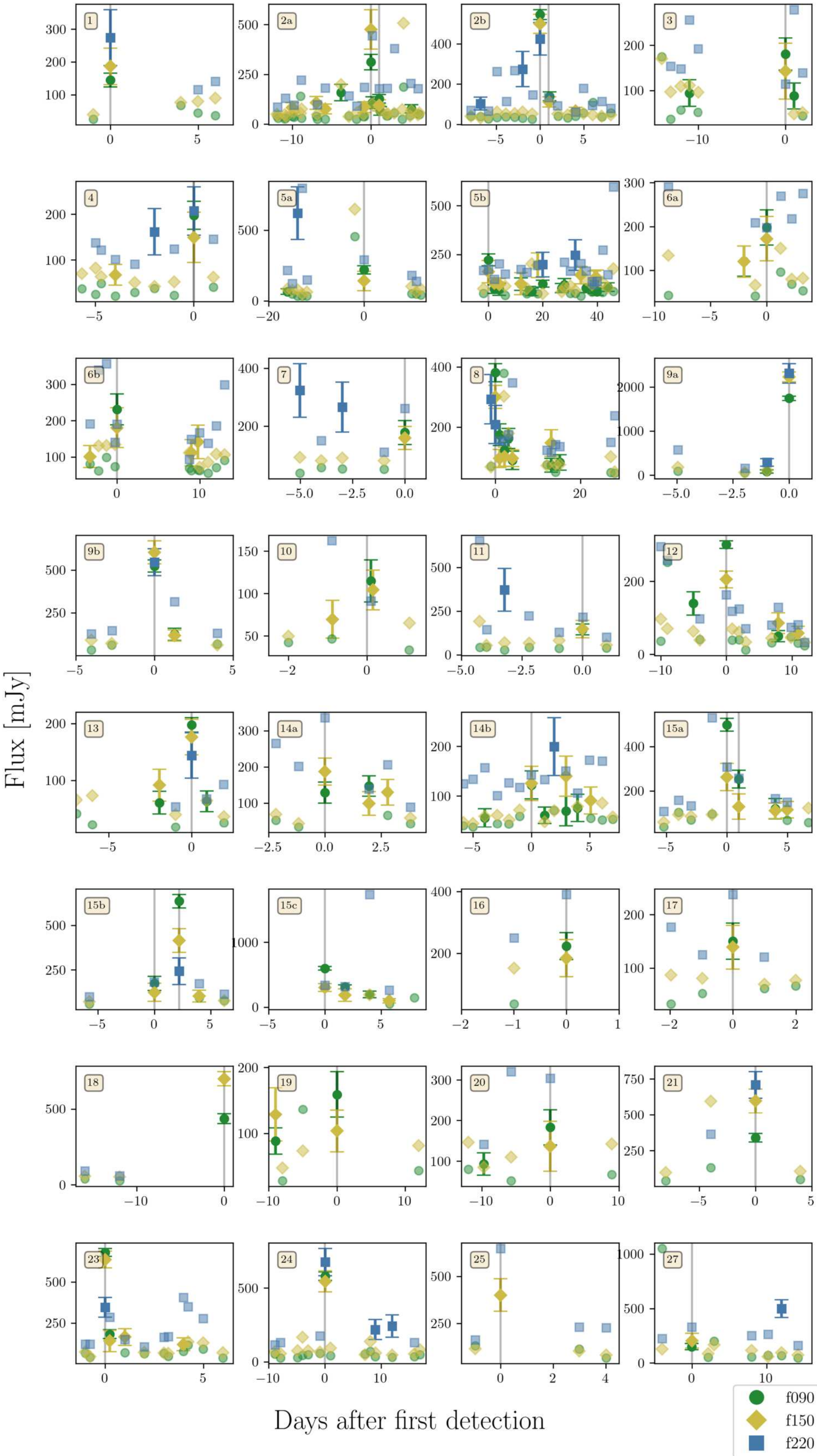

**Figure A1.** Flux light curves from the depth-1 maps. The flux is measured by detecting a source with an S/N of at least 3 and within at least 2 arcminutes of the candidate's position. The data are binned in 0.5 day bins for clarity. Points with error bars indicate a detection of S/N > 3, and points without error bars show the upper flux limits for each frequency. The time of detection from the pipeline is marked by the gray vertical lines. Note the light curves for events 22 and 26 are shown in Figure A3 because they are not associated with stellar flares.

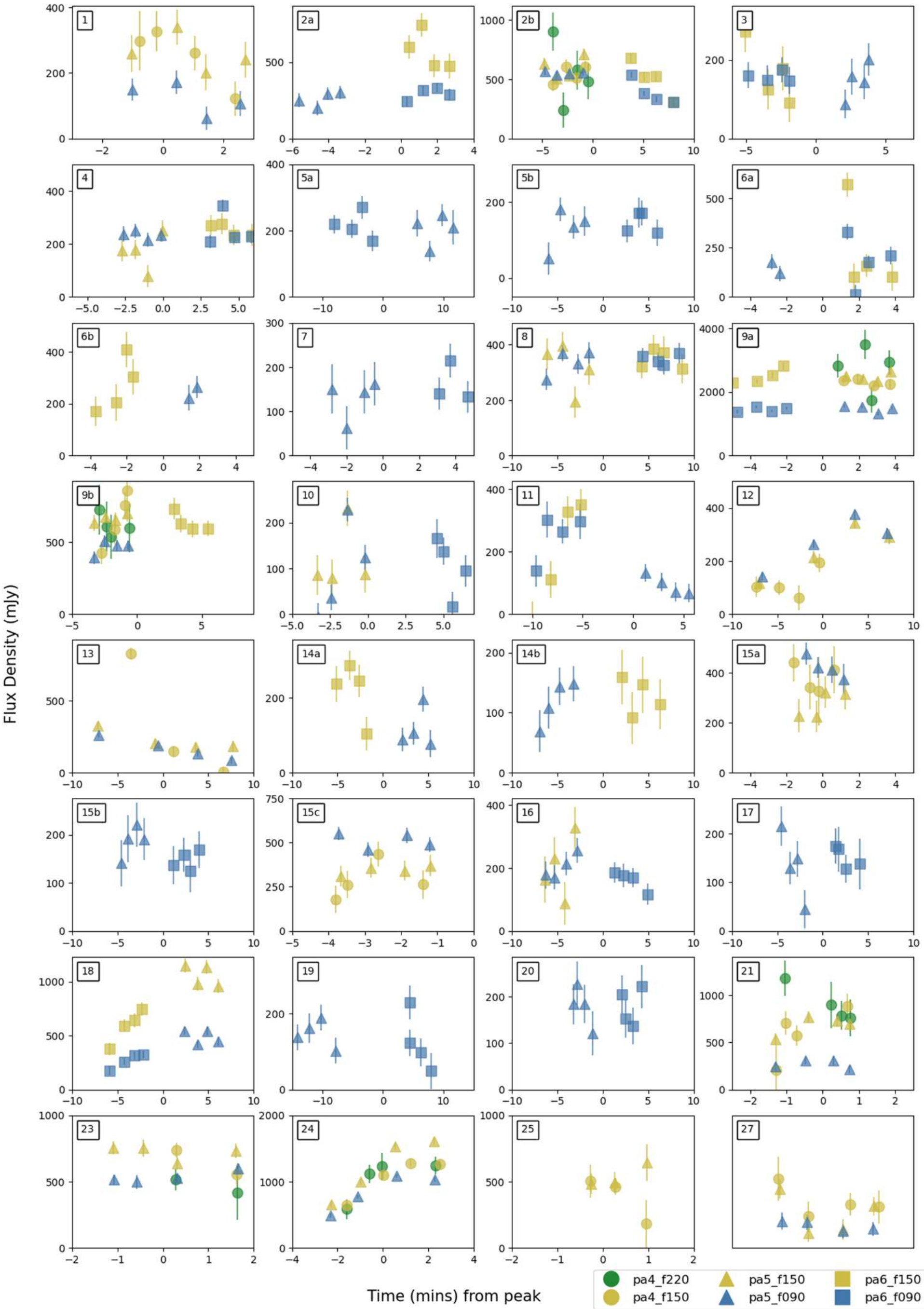

**Figure A2.** Subarray light curves from forced photometry measurements. Each array is separated into four sections, and we evaluate the flux per section as the transient event drift across the whole array.

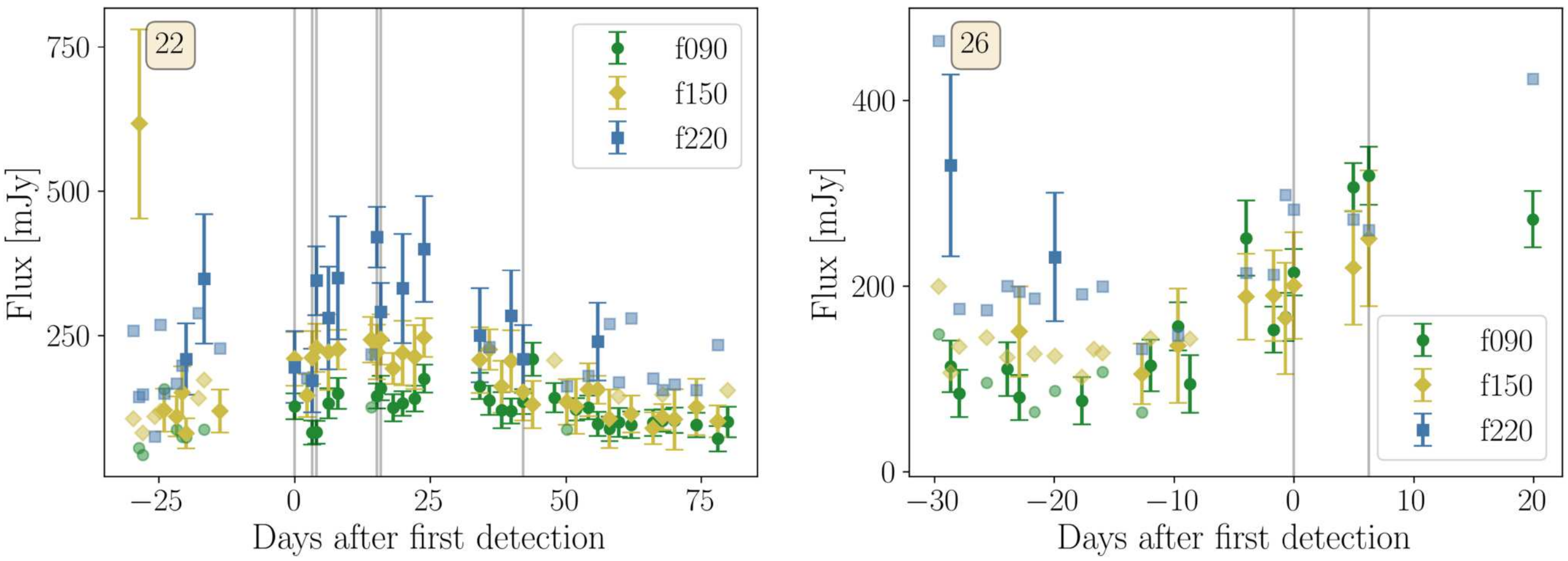

**Figure A3.** Depth-1 light curves for non-stellar-flare events. The *x*-axis measures the time since the first detection by ACT. The detections from the transient pipeline are marked with gray vertical lines. We present light curves from three frequency bands, f090, f150, and f220, shown in blue, yellow, and green, respectively. The data are binned into 0.5 day bins for clarity. Points with error bars indicate a detection of S/N> 3, and points with down arrows show upper flux limits for each frequency. Left: Event 22 is coincident with the classical nova YZ Ret. It is detected by ACT about 60 days after initial detection in X-Ray and optical bands (K. V. Sokolovsky et al. 2022). Right: Event 26 is associated with the LINER-type AGN 2MASX J19495127-363523. This AGN appears to brighten above the baseline variability.

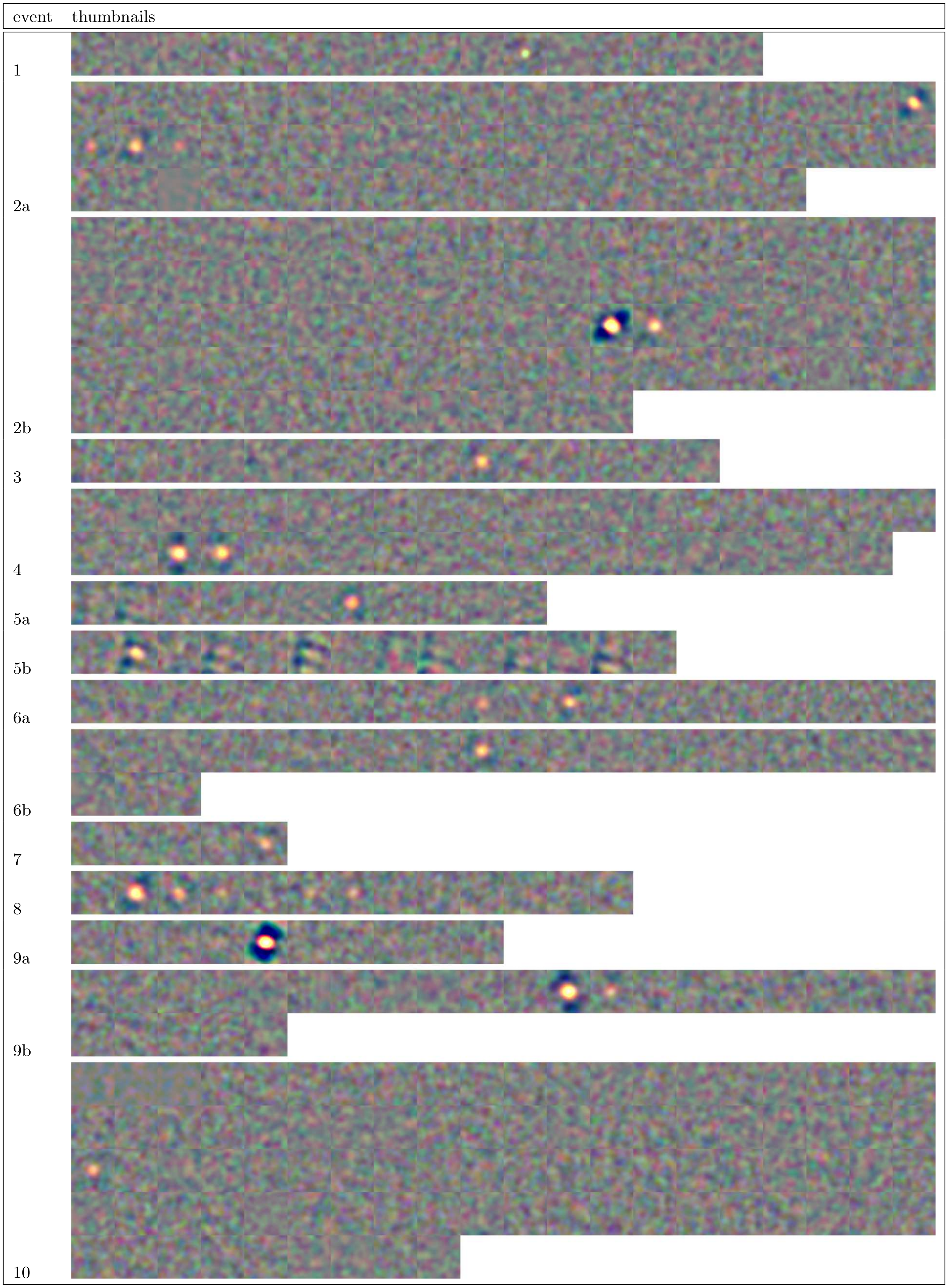

**Figure A4.** Three-color $10' \times 10'$ images of candidates from depth-1 maps about 1 month before and after detection. Red, blue, and green correspond to f090, f150, and f220, respectively. Note that the thumbnails are chronological but have inconsistent spacing in time. These plots are included for illustration rather than quantitative analysis.

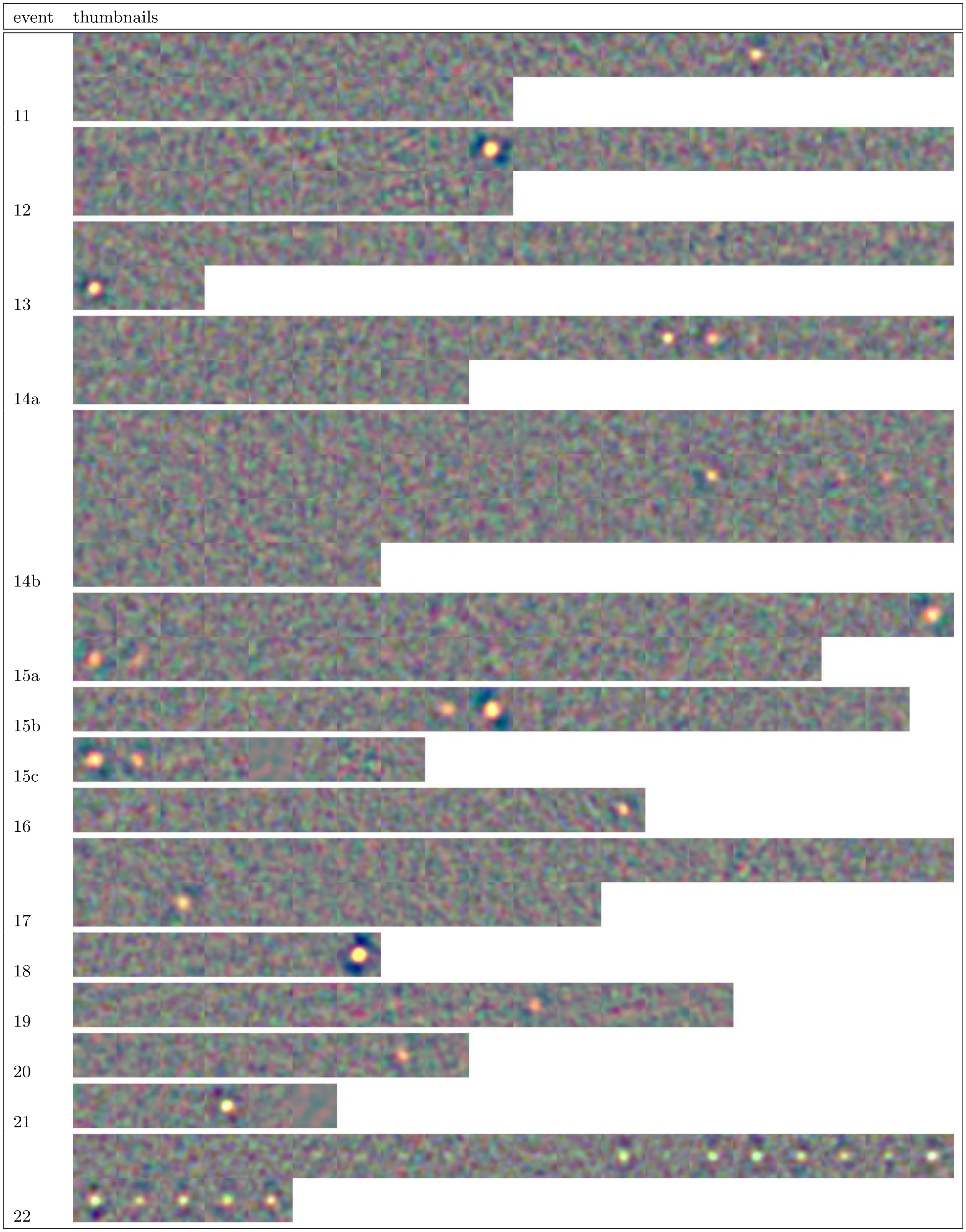

**Figure A4.** (Continued.)

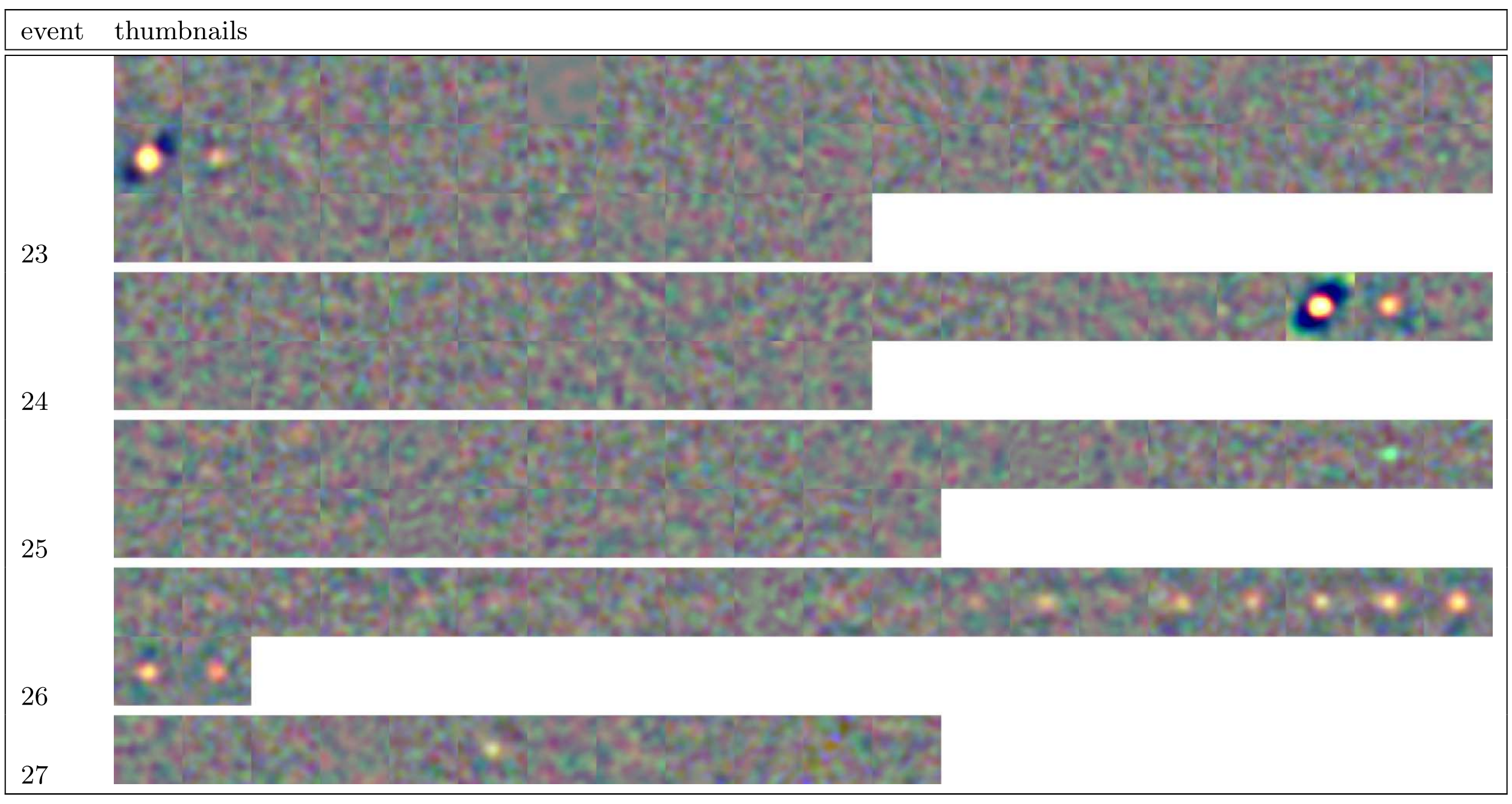

**Figure A4.** (Continued.)

**Table A1**
Catalog of the Positions of Observed Transient Events from This Work

| Ind | Name (ACT−T) | R.A. (hh:mm:ss) | Decl. (deg:arcmin:arcsec) | Pos. Acc. (arcsec) |
|---|---|---|---|---|
| 1 | J175954+104419 | 17:59:54 | +10:44:20 | 13 |
| 2a | J142555+141201 | 14:25:55 | +14:12:04 | 12 |
| 2b | J142556+141210 | 14:25:56 | +14:12:11 | 14 |
| 3 | J203622+121538 | 20:36:22 | +12:15:40 | 11 |
| 4 | J224500−331529 | 22:44:60 | -33:15:32 | 12 |
| 5a | J225303+165023 | 22:53:02 | +16:50:24 | 4 |
| 5b | J225302+165012 | 22:53:03 | +16:50:13 | 20 |
| 6a | J051922−072036 | 05:19:22 | -07:20:35 | 8 |
| 6b | J051921−072041 | 05:19:21 | -07:20:42 | 15 |
| 7 | J011636−022950 | 01:16:36 | -02:29:49 | 21 |
| 8 | J200800+160955 | 20:07:59 | +16:09:54 | 9 |
| 9a | J192832−350757 | 19:28:32 | -35:07:55 | 6 |
| 9b | J192832−350757 | 19:28:32 | -35:07:55 | 7 |
| 10 | J130046+122233 | 13:00:46 | +12:22:34 | 11 |
| 11 | J050048−571534 | 05:00:48 | -57:15:36 | 16 |
| 12 | J085814+194546 | 08:58:14 | +19:45:47 | 3 |
| 13 | J101936+195218 | 10:19:36 | +19:52:19 | 14 |
| 14a | J130530+124936 | 13:05:30 | +12:49:37 | 7 |
| 14b | J130529+124937 | 13:05:29 | +12:49:37 | 13 |
| 15a | J033647+003511 | 03:36:47 | +00:35:13 | 6 |
| 15b | J033647+003512 | 03:36:48 | +00:35:20 | 12 |
| 15c | J033647+003510 | 03:36:47 | +00:35:10 | 16 |
| 16 | J125045+113329 | 12:50:45 | +11:33:29 | 14 |
| 17 | J174147+022842 | 17:41:47 | +02:28:41 | 31 |
| 18 | J181516−492748 | 18:15:16 | -49:27:47 | 7 |
| 19 | J180723+194223 | 18:07:23 | +19:42:22 | 16 |
| 20 | J040942−075325 | 04:09:42 | -07:53:24 | 11 |
| 21 | J193939−060344 | 19:39:38 | -06:03:43 | 10 |
| 22 | J035830−544640 | 03:58:30 | -54:46:37 | 18 |
| 23 | J165122−005006 | 16:51:22 | -00:50:06 | 8 |
| 24 | J173353+165508 | 17:33:53 | +16:55:08 | 25 |
| 25 | J204747−363544 | 20:47:47 | -36:35:46 | 17 |
| 26 | J194951−363523 | 19:49:52 | -36:35:20 | 18 |
| 27 | J001309+053532 | 00:13:09 | +05:35:31 | 16 |

**Note.** Each unique number identifier is associated with a different location on the sky. Transients marked with a letter have multiple events at the same position in the sky. Each event's position error is evaluated as the standard deviation of the coordinates observed by different array–frequency combinations.

**Table A2**
Flare Properties for All Transient Events Detected in This Work

| Ind | Peak Flux (mJy) | | | Mean Flux (mJy) | | | Time | | | $\alpha$ |
|---|---|---|---|---|---|---|---|---|---|---|
| | f220 | f150 | f090 | f220 | f150 | f090 | Peak (UTC) | Rise (hr) | Fall (hr) | |
| 1 | 274 ± 86 | 186 ± 56 | 145 ± 21 | 5 ± 3 | 1 ± 1 | 6 ± 2 | 2017-08-05 03:26:35 | <24 | <48 | 0.7 ± 0.4 |
| 2a | ⋯ | 476 ± 100 | 312 ± 39 | −2 ± 2 | 2 ± 1 | 6 ± 1 | 2017-09-17 16:54:14 | >19 | >24 | 1.0 ± 0.6 |
| 2b | 424 ± 81 | 499 ± 47 | 545 ± 25 | −1 ± 2 | 2 ± 1 | 5 ± 1 | 2018-11-21 13:09:28 | ≫0.08 | >24 | −0.3 ± 0.2 |
| 3 | ⋯ | 143 ± 62 | 181 ± 36 | −1 ± 3 | 3 ± 1 | 5 ± 2 | 2017-10-08 02:26:16 | ≫0.07 | >24 | −0.5 ± 1.1 |
| 4 | 207 ± 52 | 182 ± 40 | 240 ± 22 | 5 ± 2 | 4 ± 1 | 9 ± 1 | 2017-10-08 22:19:15 | ≫0.07 | >23 | −0.3 ± 0.3 |
| 5a | ⋯ | 139 ± 69 | 218 ± 29 | 5 ± 3 | 8 ± 1 | 19 ± 2 | 2017-11-15 0:47:40 | ≫0.08 | ≫0.08 | −1.0 ± 1.2 |
| 5b | ⋯ | 161 ± 57 | 222 ± 31 | 4 ± 3 | 7 ± 1 | 21 ± 2 | 2019-06-06 08:22:39 | ∼0.08 | ≫0.08 | −0.7 ± 0.9 |
| 6a | ⋯ | 172 ± 51 | 198 ± 40 | 8 ± 4 | 4 ± 2 | 13 ± 3 | 2017-11-24 02:53:11 | >120 | <48 | −0.3 ± 0.8 |
| 6b | ⋯ | 181 ± 55 | 231 ± 42 | 8 ± 4 | 4 ± 2 | 14 ± 3 | 2018-10-21 10:48:30 | >0.05 | <192 | −0.6 ± 0.8 |
| 7 | ⋯ | 159 ± 40 | 178 ± 42 | 0 ± 5 | 0 ± 2 | −2 ± 3 | 2018-05-31 10:37:35 | <48 | ≫0.07 | −0.3 ± 0.8 |
| 8 | 208 ± 64 | 300 ± 38 | 381 ± 30 | −3 ± 3 | 3 ± 1 | 15 ± 2 | 2018-09-11 23:43:42 | ≫0.1 | >24 | −0.6 ± 0.3 |
| 9a | 2307 ± 222 | 2228 ± 109 | 174 1± 48 | 5 ± 4 | 20 ± 1 | 38 ± 3 | 2018-10-04 02:19:22 | >24 | ≫0.07 | 0.4 ± 0.1 |
| 9b | 545 ± 79 | 604 ± 69 | 523 ± 35 | 5 ± 4 | 12 ± 1 | 38 ± 3 | 2019-08-10 23:21:59 | ≫0.06 | >24 | 0.1 ± 0.2 |
| 10 | ⋯ | 104 ± 23 | 115 ± 25 | −1 ± 2 | 0 ± 5 | 1 ± 1 | 2018-10-25 12:50:17 | ∼0.08 | <19 | −0.2 ± 0.7 |
| 11 | ⋯ | 148 ± 50 | 145 ± 31 | 0 ± 3 | 2 ± 1 | 0 ± 2 | 2018-11-08 09:06:50 | ∼0.08 | >0.17 | 0.0 ± 0.9 |
| 12 | ⋯ | 206 ± 23 | 301 ± 11 | 3 ±2 | 0 ± 1 | 2 ± 1 | 2018-11-15 10:21:42 | ∼0.08 | ≫0.10 | −0.9 ± 0.3 |
| 13 | 144 ± 40 | 176 ± 31 | 198 ± 13 | 0 ±2 | 0 ± 1 | 4 ± 1 | 2018-12-21 09:19:05 | >24 | >21 | −0.3 ± 0.3 |
| 14a | ⋯ | 187 ± 37 | 129 ± 29 | 5 ±2 | 1 ± 1 | 6 ± 1 | 2019-03-13 08:38:52 | <48 | >144 | 0.9 ± 0.7 |
| 14b | ⋯ | 126 ± 34 | 122 ± 28 | 4 ±2 | 1 ± 1 | 6 ± 1 | 2019-08-02 19:18:16 | >0.08 | >120 | 0.1 ± 0.8 |
| 15a | ⋯ | 262 ± 62 | 498 ± 29 | 1 1±4 | 12 ± 1 | 32 ± 3 | 2019-08-10 13:35:32 | <48 | >96 | −1.5 ± 0.6 |
| 15b | ⋯ | 124 ± 51 | 174 ± 40 | 10 ±4 | 11 ± 1 | 30 ± 3 | 2019-09-26 05:11:56 | >48 | ≫0.07 | −0.8 ± 1.1 |
| 15 c | ⋯ | 304 ± 61 | 599 ± 27 | 12 ±4 | 12 ± 1 | 32 ± 3 | 2021-09-29 10:19:17 | ≫0.07 | >120 | −1.6 ± 0.5 |
| 16 | ⋯ | 184 ± 60 | 223 ± 43 | −3 ±2 | 2 ± 1 | 5 ± 1 | 2019-09-17 15:39:05 | >0.07 | >0.07 | −0.4 ± 0.9 |
| 17 | ⋯ | 139 ± 41 | 151 ± 34 | −1 ±4 | 2 ± 1 | 5 ± 3 | 2019-10-26 16:57:32 | <23 | <23 | −0.2 ± 0.9 |
| 18 | ⋯ | 699 ± 47 | 435 ± 32 | −2 ±6 | 6 ± 2 | 16 ± 5 | 2019-11-08 22:27:58 | ∼0.17 | ≫0.07 | 1.1 ± 0.2 |
| 19 | ⋯ | 104 ± 32 | 159 ± 35 | 0 ± 2 | 3 ± 1 | 5 ± 2 | 2019-11-13 18:12:29 | <125 | <288 | −1.0 ± 0.9 |
| 20 | ⋯ | 136 ± 61 | 183 ± 43 | 3 ± 4 | 4 ± 2 | 18 ± 3 | 2019-12-07 00:41:54 | ≫0.07 | ≫0.07 | −0.7 ± 1.2 |
| 21 | 709 ± 93 | 597 ± 83 | 340 ± 30 | 17 ± 6 | 3 ± 2 | 9 ± 4 | 2020-06-18 09:20:21 | <125 | <120 | 0.9 ± 0.2 |
| 22 | 195 ± 62 | 210 ± 46 | 128 ± 22 | 13 ± 3 | 12 ± 1 | 18 ± 2 | 2020-09-05 05:55:29 | >240 | ∼600 | 0.6 ± 0.4 |
| 23 | 346 ± 60 | 637 ± 52 | 680 ± 25 | −1 ± 3 | 2 ± 1 | 11 ± 2 | 2020-09-06 19:44:50 | <18 | >23 | −0.6 ± 0.1 |
| 24 | 676 ± 93 | 1005 ± 52 | 956 ± 22 | −1 ± 2 | 1 ± 1 | 0 ± 2 | 2020-10-17 18:47:17 | ∼0.10 | >3 | −0.1 ± 0.1 |
| 25 | ⋯ | 401 ± 87 | ⋯ | 2 ± 4 | 3 ± 1 | 5 ± 3 | 2020-11-13 17:59:00 | <24 | <72 | ⋯ |
| 26 | ⋯ | 201 ± 57 | 215 ± 25 | 13 ± 4 | 22 ± 1 | 50 ± 3 | 2021-05-19 05:09:51 | ∼480 | ⋯ | −0.2 ± 0.7 |
| 27 | ⋯ | 205 ± 68 | 151 ± 28 | 2 ± 4 | 0 ± 1 | −2 ± 3 | 2022-05-21 10:00:08 | <96 | <72 | 0.7 ± 0.9 |

**Note.** We provide the peak fluxes of each frequency. If a source is detected by more than one array in a given frequency band, we cite a weighted average. Where there is not a $>5\sigma$ S/N detection, we do not report a peak flux. We calculate the mean flux at each band by applying a matched filter to the mean sky map using data from 2017 to 2021 and measuring a weighted average of the flux at the transient's position, not including detection times or maps with known pointing errors. The peak time is defined as the time of the maximum flux in the f090 frequency band. The rise and fall times are evaluated from the candidates' light curves and subarray light curves (see Figure A2). The "≫" sign of the candidate's rise and fall time means that we do not see a minute-wise flux density change within the peak scan but there is a $>5$ day time gap between the peak scan and the adjacent scan. The spectral index $\alpha$ is evaluated by fitting a power law of $S_\nu \propto \nu^\alpha$ of frequency $\nu$ using peak flux density $S_\nu$. Note that Event 25 does not have a spectral index calculation because the peak flux is measured in only one frequency band.

**Table A3**
Column Descriptions for the Machine-readable File "ACT Depth-1 Transient Catalog" Accompanying This Paper

| Unit | Label | Explanation |
|---|---|---|
| ⋯ | Name | ACT-T Transient name corresponding to ACT location. |
| ⋯ | Seq | Numerical ID corresponding to ID in this paper. |
| deg | RAdeg | R.A. in degrees of the transient as measured in the ACT depth-1 map. |
| deg | DEdeg | Decl. in degrees of the transient as measured in the ACT depth-1 map. |
| deg | PosErr | Position error in degrees evaluated as the variance of coordinates observed by all $>5\sigma$ detections in all depth-1 maps and array–frequency combinations. |
| mJy | f220-Peak | Peak flux density in millijansky of the depth-1 light curve for the f220 band. |
| mJy | f150-Peak | Peak flux density in millijansky of the depth-1 light curve for the f150 band. |
| mJy | f090-Peak | Peak flux density in millijansky of the depth-1 light curve for the f090 band. |
| mJy | e_f220-Peak | Error on the peak flux density in millijansky for the f220 band. |
| mJy | e_f150-Peak | Error on the peak flux density in millijansky for the f150 band. |
| mJy | e_f090-Peak | Error on the peak flux density in millijansky for the f090 band. |
| mJy | f220-Mean | Mean flux density in millijansky for the f220 band across all depth-1 maps with transient detections omitted. |
| mJy | f150-Mean | Mean flux density in millijansky for the f150 band across all depth-1 maps with transient detections omitted. |
| mJy | f090-Mean | Mean flux density in millijansky for the f090 band across all depth-1 maps with transient detections omitted. |
| mJy | e_f220-Mean | Mean flux density error in millijansky for the f220 band. |
| mJy | e_f150-Mean | Mean flux density error in millijansky for the f150 band. |
| mJy | e_f090-Mean | Mean flux density error in millijansky for the f090 band. |
| d | f220-tpeak | MJD of peak flux density for the f220 band. |
| d | f150-tpeak | MJD of peak flux density for the f150 band. |
| d | f090-tpeak | MJD of peak flux density for the f090 band. |
| ⋯ | Sp+Index | Spectral index calculated at peak flux density by fitting a power law. |
| ⋯ | e_Sp+Index | Spectral index error. |
| ⋯ | Simbad-Id | SIMBAD ID of most likely association. |
| ⋯ | Type | Type of star given by SIMBAD. |
| ⋯ | SpType | Spectral type given by SIMBAD. |
| ⋯ | Sep | Separation between ACT observation and Gaia coordinates of associated SIMBAD object. |
| ⋯ | pval | Chance of a random association given optical magnitude and sky separation of the counterpart. |
| pc | Dist | Distance in parseconds to the counterpart. |

(This table is available in its entirety in machine-readable form in the online article.)

**Table A4**
Transient Counterparts from the SIMBAD Database

| Name | SIMBAD ID | Type | Mag. | Pos. Err. (arcsec) | Sep. (arcsec) | Chance | Dist. (pc) |
|---|---|---|---|---|---|---|---|
| 1 | 1SWASP J175954.36+104418.9 | BYDraV$^*$ | 10.75 | 12.65 | 3.26 | 8.66e−05 | 307.78 |
| 2a/b | StKM 1−1155 | Low-Mass$^*$ (M0.0Ve) | 10.91 | 12.10 | 10.99 | 2.94e−04 | 157.92 |
| 3 | ASAS J203622+1215.3 | BYDraV$^*$ | 9.28 | 10.56 | 3.10 | 2.09e−05 | 488.40 |
| 4 | V$^*$ TX PsA | Eruptive$^*$ (M5IVe) | 11.84 | 12.14 | 4.01 | 1.08e−04 | 20.83 |
| 5a/b | V$^*$ IM Peg | RSCVnV$^*$ (K2III) | 5.66 | 3.59 | 6.81 | 8.95e−07 | 98.37 |
| 6a/b | HD 34736 | PulsV$^*$ (ApSi) | 7.79 | 8.21 | 18.17 | 1.46e−04 | 372.44 |
| 7 | $^*$ 39 Cet | RSCVnV$^*$ (G6III:eFe-2) | 5.24 | 21.02 | 12.21 | 9.59e−07 | 74.84 |
| 8 | HD 191179 | SB$^*$ (G5) | 7.96 | 9.08 | 2.62 | 6.10e−06 | 219.46 |
| 9a/b | HD 182928 | RotV$^*$ (G8IIIe) | 9.37 | 6.18 | 2.41 | 7.28e−06 | 196.58 |
| 10 | BD+13 2618 | Eruptive$^*$ (M0V) | 8.91 | 11.35 | 2.01 | 1.56e−06 | 11.51 |
| 11 | CD−57 1054 | Eruptive$^*$ (M0Ve) | 9.35 | 16.24 | 11.07 | 1.58e−04 | 26.87 |
| 12 | G 9−38B | HighPM$^*$ (M7V) | 12.49 | 3.47 | 3.42 | 1.54e−04 | 5.10 |
| 13 | V$^*$ AD Leo | Eruptive$^*$ (dM3) | 8.21 | 13.73 | 7.19 | 1.20e−05 | 4.97 |
| 14a/b | HD 113714 | RSCVnV$^*$ (G4V) | 9.87 | 7.17 | 7.60 | 5.47e−05 | 316.38 |
| 15a/b/c | HD 22468 | RSCVnV$^*$ (K2:Vnk) | 5.60 | 6.03 | 8.71 | 1.46e−06 | 29.43 |
| 16 | UCAC4 508−055499 | RSCVnV$^*$ | 12.43 | 13.91 | 8.52 | 5.22e−04 | 420.27 |
| 17 | BD+02 3384 | RotV$^*$ (G9III) | 9.09 | 30.70 | 16.81 | 3.05e−04 | 395.01 |
| 18 | 2MASS J18151564−4927472 | HighPM$^*$ (M3) | 11.72 | 7.47 | 2.90 | 2.55e−04 | 61.99 |
| 19 | HD 347929 | RotV$^*$ (K0) | 9.04 | 15.87 | 9.04 | 1.12e−04 | 240.74 |
| 20 | V$^*$ EI Eri | RSCVnV$^*$ (G3V) | 6.95 | 11.19 | 17.25 | 6.31e−05 | 54.41 |
| 21 | HD 185510 | RSCVnV$^*$ (K0III+sdB) | 7.81 | 9.82 | 7.47 | 1.62e−05 | 178.58 |
| 22 | V$^*$ YZ Ret | Nova (sdB?/DA?) | 16.26 | 18.22 | 1.49 | 3.08e−04 | 2644.80 |
| 23 | V$^*$ V2700 Oph | BYDraV$^*$ | 11.21 | 7.66 | 4.27 | 1.12e−04 | 312.81 |
| 24 | V$^*$ V1274 Her | BYDraV$^*$ (M6V) | 12.40 | 24.77 | 3.00 | 2.31e−04 | 16.42 |
| 25 | V$^*$ BO Mic | BYDraV$^*$ (K3V(e)) | 8.93 | 17.26 | 22.27 | 4.11e−04 | 51.02 |
| 26 | 2MASX J19495127−3635239 | LINER | … | 17.98 | 6.85 | … | … |
| 27 | V$^*$ DV Psc | SB$^*$ (K5Ve) | 10.21 | 16.39 | 7.79 | 1.10e−04 | 42.16 |

**Note.** For repeating events, we take the lower position error to calculate the chance of a random association. For each independent transient position we cite the most likely counterpart from SIMBAD. In the case of binary objects, or systems with more than one star, we cite the object with the lower chance of a random association with the transient candidate. The separations are calculated using Gaia positions for all but Event 26 where we use the position from SIMBAD. For all events except Event 7 we calculate a density of Gaia stars with the magnitude of the candidate or brighter to determine the chance of a random association using Equation (3) (*p*-value). For Event 7 this is calculated using a 6° radius due to a low density of stars with this magnitude or brighter in this region. We also include the spectral type, if known, in the type column.

**Table A5**
Column Descriptions for the Depth-1 Light-curve Machine-readable Files

| Units | Label | Explanation |
|---|---|---|
| Hz | Freq | Frequency of observation assuming the CMB blackbody spectrum. |
| … | Array | Map array (pa4, pa5, or pa6). |
| deg | RAdeg | R.A. of the observation. If there is a $>3\sigma$ detection, this is the center of mass by flux. Otherwise, this is the position of the event. |
| deg | DEdeg | Decl. of the observation. If there is a $>3\sigma$ detection, this is the center of mass by flux. Otherwise, this is the position of the event. |
| d | MJD | The MJD of the observation, read-off from the map. |
| mJy | Flux | The flux of the observation, read-off from the map. |
| mJy | e_Flux | The error on the flux, read-off from the map. |
| mJy | E_Flux | If there is no detection we calculate the flux that captures 95% of the probability assuming a Gaussian measurement. This serves as an upper flux limit. |

**Note.** The files are named by each event's index and includes all available depth-1 map data from pa4, pa5, and pa6. Only one file is given for multiple events at the same location.

(This table is available in its entirety in machine-readable form in the online article.)

## ORCID iDs

Emily Biermann https://orcid.org/0000-0002-2840-9794
Yaqiong Li https://orcid.org/0000-0001-8093-2534
Sigurd Naess https://orcid.org/0000-0002-4478-7111
Steve K. Choi https://orcid.org/0000-0002-9113-7058
Susan E. Clark https://orcid.org/0000-0002-7633-3376
Mark Devlin https://orcid.org/0000-0002-3169-9761
Jo Dunkley https://orcid.org/0000-0002-7450-2586
P. A. Gallardo https://orcid.org/0000-0001-9731-3617
Yilun Guan https://orcid.org/0000-0002-1697-3080
Allen Foster https://orcid.org/0000-0002-7145-1824
Matthew Hasselfield https://orcid.org/0000-0002-2408-9201
Carlos Hervías-Caimapo https://orcid.org/0000-0002-4765-3426

Matt Hilton https://orcid.org/0000-0002-8490-8117
Adam D. Hincks https://orcid.org/0000-0003-1690-6678
Anna Y. Q. Ho https://orcid.org/0000-0002-9017-3567
John C. Hood, II https://orcid.org/0000-0003-4157-4185
Kevin M. Huffenberger https://orcid.org/0000-0001-7109-0099
Arthur Kosowsky https://orcid.org/0000-0002-3734-331X
Michael D. Niemack https://orcid.org/0000-0001-7125-3580
John Orlowski-Scherer https://orcid.org/0000-0003-1842-8104
Lyman Page https://orcid.org/0000-0002-9828-3525
Bruce Partridge https://orcid.org/0000-0001-6541-9265
Maria Salatino https://orcid.org/0000-0003-4006-1134
Cristóbal Sifón https://orcid.org/0000-0002-8149-1352
Suzanne T. Staggs https://orcid.org/0000-0002-7020-7301
Cristian Vargas https://orcid.org/0000-0001-5327-1400
Edward J. Wollack https://orcid.org/0000-0002-7567-4451